\pdfoutput=1

\documentclass[aps,prd,twocolumn,preprintnumbers,superscriptaddress,nofootinbib,floatfix]{revtex4-1}

\usepackage{amsmath,amssymb}
\usepackage{bm}
\usepackage{graphicx}
\usepackage{epstopdf}
\usepackage{hyperref}
\usepackage{array}
\usepackage[utf8]{inputenc}
\usepackage{soul}
\usepackage[usenames, dvipsnames]{color}

\usepackage{subfigure}
\usepackage{slashed}
\usepackage{afterpage}
\usepackage{psfrag}
\usepackage{framed}
\usepackage{multirow}

\usepackage[shortlabels]{enumitem}

\usepackage{array}
\newcolumntype{L}[1]{>{\raggedright\let\newline\\\arraybackslash\hspace{0pt}}m{#1}}
\newcolumntype{C}[1]{>{\centering\let\newline\\\arraybackslash\hspace{0pt}}m{#1}}
\newcolumntype{R}[1]{>{\raggedleft\let\newline\\\arraybackslash\hspace{0pt}}m{#1}}

\enlargethispage{\baselineskip}

\hypersetup{
 colorlinks = true,
 citecolor = blue
}

\newcommand{\ccBor}{Na$_2$(B$_4$O$_5$)(OH)$_4\!\cdot\!8($H$_2$O)}
\newcommand{\ccCat}{Mg$_{2.92}$Fe$_ {0.01}$(PO$_4$)$_{2.01}\!\cdot\!22.05$(H$_2$O)}
\newcommand{\ccEps}{Mg(SO$_4)\!\cdot\!7($H$_2$O)}
\newcommand{\ccGyp}{Ca(SO$_4)\!\cdot\!2$(H$_2$O)}
\newcommand{\ccHal}{NaCl}
\newcommand{\ccMir}{Na$_2($SO$_4)\!\cdot\!10($H$_2$O)}

\newcommand{\ccBad}{ZrO$_2$}
\newcommand{\ccNch}{Mn$^{2+}_2$SiO$_3$(OH)$_2 \! \cdot \! ($H$_2$O)}
\newcommand{\ccNic}{NiCl$_2\!\cdot\!6($H$_2$O)}
\newcommand{\ccOli}{Mg$_{1.6}$Fe$^{2+}_{0.4}$(SiO$_4$)}
\newcommand{\ccPhl}{KMg$_3$AlSi$_3$O$_{10}$F(OH)}

\begin{document}

\title{Paleo-detectors: Searching for Dark Matter with Ancient Minerals} 

\newcommand{\OKC}{\affiliation{The Oskar Klein Centre for Cosmoparticle Physics, Department of Physics, Stockholm University, Alba Nova, 10691 Stockholm, Sweden}}
\newcommand{\Nordita}{\affiliation{Nordita, KTH Royal Institute of Technology and Stockholm University, Roslagstullsbacken 23, 10691 Stockholm, Sweden}}
\newcommand{\LCTP}{\affiliation{Leinweber Center for Theoretical Physics, University of Michigan, Ann Arbor, MI 48109, USA}}
\newcommand{\NCBJ}{\affiliation{National Centre for Nuclear Research, 05-400 Otwock, \'{S}wierk, Poland}}

\author{Andrzej~K.~Drukier}
\email{adrukier@gmail.com}
\OKC

\author{Sebastian~Baum}
\email{sbaum@fysik.su.se}
\OKC
\Nordita

\author{Katherine~Freese}
\email{ktfreese@umich.edu}
\OKC
\Nordita
\LCTP

\author{Maciej~G\'{o}rski}
\email{maciej.gorski@ncbj.gov.pl}
\NCBJ

\author{Patrick~Stengel}
\email{patrick.stengel@fysik.su.se}
\OKC

\preprint{NORDITA-2018-117}
\preprint{LCTP-18-25}

\begin{abstract}
We explore {\it paleo-detectors} as an approach to the direct detection of Weakly Interacting Massive Particle (WIMP) dark matter radically different from conventional detectors.  Instead of instrumenting a (large) target mass in a laboratory in order to observe WIMP-induced nuclear recoils in real time, the approach is to examine ancient minerals for traces of WIMP-nucleus interactions recorded over timescales as large as 1\,Gyr.  Here, we discuss the paleo-detector proposal in detail, including background sources and possible target materials. In order to suppress backgrounds induced by radioactive contaminants such as uranium, we propose to use minerals found in marine evaporites or in ultra-basic rocks. We estimate the sensitivity of paleo-detectors to spin-independent and spin-dependent WIMP-nucleus interactions. The sensitivity to low-mass WIMPs with masses $m_\chi \lesssim 10\,$GeV extends to WIMP-nucleon cross sections many orders of magnitude smaller than current upper limits. For heavier WIMPs with masses $m_\chi \gtrsim 30\,$GeV cross sections a factor of a few to $\sim 100$ smaller than current upper limits can be probed by paleo-detectors.
\end{abstract}

\maketitle

%*************************
\section{Introduction}
%*************************
Following our recent work~\cite{Baum:2018tfw}, we explore paleo-detectors as a radical alternative to conventional direct detection techniques for Dark Matter (DM).  Current strategies for the direct detection of Weakly Interacting Massive Particle (WIMP) DM employ large target masses, which are instrumented to observe nuclear recoils in real time~\cite{Drukier:1983gj,Goodman:1984dc,Drukier:1986tm}. Such experiments have set impressive upper limits on the strength of interactions of WIMPs with atomic nuclei~\cite{Angloher:2015ewa,Amole:2017dex,Akerib:2017kat,Aprile:2017iyp,Agnese:2017jvy,Cui:2017nnn,Petricca:2017zdp,Aprile:2018dbl}, but so far failed to report conclusive evidence for DM. The long-standing exception is the DAMA/LIBRA experiment~\cite{Bernabei:2008yi,Bernabei:2013xsa,Bernabei:2018yyw}, reporting evidence for an annual modulation signal~\cite{Drukier:1986tm,Spergel:1987kx,Freese:2012xd} compatible with DM for more than a decade~\cite{Gondolo:2005hh,Petriello:2008jj,Chang:2008xa,Fairbairn:2008gz,Savage:2008er,Bottino:2008mf,Baum:2018ekm,Kang:2018qvz,Herrero-Garcia:2018lga}. However, the claimed signal is in tension with null-results from other direct detection experiments~\cite{Savage:2008er,Savage:2009mk,Savage:2010tg,McCabe:2011sr,Catena:2016hoj}.

Currently, there are two major trends in the evolution of direct detection: experiments primarily aimed at WIMPs with masses $m_\chi \gtrsim 15\,$GeV plan to utilize larger target masses, approaching the 100\,t scale~\cite{Aprile:2015uzo,Mount:2017qzi,Aalbers:2016jon,Aalseth:2017fik,Amaudruz:2017ibl}. Such experiments use liquid noble gas targets and are sensitive to nuclear recoils with energies larger than $\mathcal{O}(1)\,$keV. For lighter WIMPs with masses $m_\chi \lesssim 15\,$GeV, the main challenge is to observe nuclear recoils with small energies. Experiments using new approaches have been proposed, including cryogenic bolometric detectors that aim for target masses of a few kg and recoil energy thresholds of $\mathcal{O}(100)\,$eV~\cite{Angloher:2015eza,Agnese:2016cpb}. 

Further, there is a major effort~\cite{Daw:2011wq,Battat:2014van,Riffard:2013psa,Santos:2013hpa,Monroe:2012qma,Leyton:2016nit,Miuchi:2010hn,Nakamura:2015iza, Battat:2016xxe} to develop directional detectors~\cite{Spergel:1987kx}, where the ability to determine the direction of the nuclear recoil would allow for powerful background rejection. Other recent proposals for direct detection include nm-scale detectors~\cite{Drukier:2012hj,Drukier:2017piu} and concepts using molecular biology~\cite{Lopez:2014oea,Drukier:2014rea}.

Paleo-detectors are radically different from conventional direct detection experiments: Instead of instrumenting a (large) target mass in a laboratory in order to observe WIMP-induced nuclear recoils in real time, we propose to examine ancient minerals for traces of WIMP-nucleus interactions recorded over timescales as large as 1\,Gyr~\cite{Baum:2018tfw}. Recoiling nuclei leave damage tracks in certain classes of minerals, so-called solid state track detectors (SSTDs). Once created, these tracks are preserved over time-scales larger than a billion years. WIMP-induced recoils would leave damage tracks with lengths up to $\mathcal{O}(500)\,$nm in typical natural targets. Hence, in paleo-detectors the challenge is to reconstruct such nano-scale features in natural minerals instead of an $\mathcal{O}(1)$ number of phonons, electrons, or photons as typically done in conventional direct detection experiments.

Our work on paleo-detectors builds on a long history of experiments. The idea to replace large detector masses with long exposure times goes back to searches for magnetic monopoles accumulated in ancient rocks~\cite{Goto:1958,Goto:1963zz,Fleischer:1970zy,Alvarez:1970zu,Kolm:1971xb,Eberhard:1971re,Ross:1973it,Kovalik:1986zz,Jeon:1995rf}. In a different approach, Refs.~\cite{Fleischer:1969mj,Fleischer:1970vm,Price:1983ax,Price:1986ky,Ghosh:1990ki} searched for damage tracks accumulated in ancient minerals from throughgoing monopoles; see Ref.~\cite{Fleischer383} for an early review of damage tracks in ancient minerals. In a related idea, Ref.~\cite{Collar:1999md} proposed to search for highly ionizing particles via the formation of Buckminister fullerenes (C$_{60}$). Snowden-Ifft et al. were the first to look for signatures of WIMP DM in muscovite mica~\cite{SnowdenIfft:1995ke}. These authors used atomic force microscopy to search for WIMP-induced nuclear recoil tracks after cleaving and chemical etching of such mica samples. Other early work on the subject can be found in Refs.~\cite{Collar:1994mj,Engel:1995gw,SnowdenIfft:1997hd}. 

Our proposal has several advantages compared to these earlier works. First, in the intervening decades, there has been enormous progress in nano-scale read-out technology. This potentially allows both for much larger sample volumes to be studied, i.e. larger exposures to be obtained, and better background mitigation. Second, we propose to use materials obtained from depths larger than $\sim 5\,$km to shield from cosmogenic backgrounds. Third, we explore a wide variety of target materials beyond muscovite mica. With these improvements, paleo-detectors could probe WIMP-nucleon cross sections far below the upper limits obtained in Ref.~\cite{SnowdenIfft:1995ke}.

In this paper, we give a detailed description of the paleo-detector proposal. In Sec.~\ref{sec:Sig}, we describe the calculation of the WIMP-induced spectrum of damage tracks. In particular, we show a semi-analytical calculation of the track length as a function of recoiling nucleus, target material, and recoil energy in Sec.~\ref{sec:Sig_TrackLength} and compare it with the usually employed numerical calculation.

In Sec.~\ref{sec:ReadOut} we discuss some generalities of solid state track detectors and a number of possible read-out methods. We identify two realistic, though ambitious, read-out scenarios. The first scenario is particularly geared towards searching for low-mass WIMPs with masses $m_\chi \lesssim 10\,$GeV. Using helium-ion beam microscopy, we estimate that target masses of $\mathcal{O}(10)\,$mg can be read out with $\sim1\,$nm spatial resolution. Reconstructing tracks as short as $\mathcal{O}(1)\,$nm, corresponding to $\mathcal{O}(100)\,$eV nuclear recoil energy thresholds, allows for thresholds comparable to conventional direct detection experiments employing cryogenic bolometric detectors. However, assuming a 1\,Gyr old sample, the exposure in paleo-detectors for $\mathcal{O}(10)\,$mg of target material is $\varepsilon = 0.01\,$kg\,Myr. In order to achieve the same exposure with a conventional direct detection experiment, one would need to observe a target mass of $10^3\,$kg for $10\,$yr.

The second read-out scenario is more suitable for heavier WIMPs, $m_\chi \gtrsim 10\,$GeV, sacrificing some spatial resolution for larger exposure. Using Small-Angle X-ray scattering (SAXs), spatial resolutions of 15\,nm are achievable. We estimate a feasible target mass of $\mathcal{O}(100)\,$g for paleo-detectors. The corresponding exposure for a life-time of the sample of 1\,Gyr is $\varepsilon = 100\,$kg\,Myr. The nuclear recoil energy threshold corresponding to $\mathcal{O}(10)\,$nm long tracks is of the order of keV, similar to what is achievable in conventional direct detection experiments using liquid noble gas targets. However, the required target mass to achieve $\varepsilon = 100\,$kg\,Myr with an integration time of 10\,yr would be $10^7\,$kg in conventional direct detection experiments. Note that other read-out methods exist beyond what is discussed here.

In Sec.~\ref{sec:Bkg} we discuss the most relevant background sources for DM searches with paleo-detectors. For the classes of target materials considered here, we identify (broadly speaking) two different background regimes: For low-mass WIMPs with masses $m_\chi \lesssim 10\,$GeV, the largest contribution to the background budget comes from nuclear recoils induced by coherent scattering of solar neutrinos. For heavier WIMPs, the largest background source is nuclear recoils induced by fast neutrons. Depending on the target's chemical composition, the dominant source of neutrons is either spontaneous fission of heavy radioactive contaminants, or $(\alpha,n)$-reactions of the $\alpha$-particles from the decays of the heavy contaminants with the material's nuclei. Note that cosmogenic backgrounds can be avoided for paleo-detectors by using minerals obtained from sufficient depth. Conventional direct detection experiments must be operated in accessible underground laboratories. The deepest current laboratory, {\it CJPL} located in China, has an overburden of $\sim 2.4\,$km rock. For paleo-detectors we envisage using comparatively small target volumes $\lesssim (10\,{\rm cm})^3$, which can be obtained from much larger depths. For example, target materials may be obtained from the bore-cores of ultra-deep boreholes used for oil exploration and geological R\&D. Existing boreholes have depths up to 12\,km~\cite{KREMENETSKY198611}. Cosmogenic backgrounds are exponentially suppressed with depth and will play virtually no role for target materials obtained from depths larger than $\sim 5\,$km. 

In Sec.~\ref{sec:MinOpt} we discuss which minerals are best suited as targets for paleo-detectors. In order to suppress backgrounds induced by radioactive contaminants, we propose to use minerals found in marine evaporites or in ultra-basic rocks. Such minerals have significantly lower concentrations of radioactive contaminants, e.g. uranium, than typical minerals found in the Earth's crust. For searches for WIMPs with masses $m_\chi \gtrsim 10\,$GeV where the background budget is dominated by neutrons from radioactive processes, target minerals containing hydrogen, e.g. hydrites, are particularly useful. This is because hydrogen is an effective moderator of fast neutrons, reducing the number of neutron-induced nuclear recoil events.

In Sec.~\ref{sec:Reach}, we present sensitivity projections for paleo-detectors in the benchmark read-out scenarios described above. Sensitivity projections for canonical Spin-Independent (SI) WIMP-nucleus interactions have already been presented in Ref.~\cite{Baum:2018tfw}. Here, we present results for a larger selection of target materials. In addition, we show the prospects of paleo-detectors for exploring Spin-Dependent (SD) WIMP-nucleus interactions. For ease of comparison with existing limits and future projections of conventional direct detection experiments, we present sensitivity projections in the proton-only and neutron-only coupling scenarios usually employed in the direct detection literature. Both in the SI and the two SD scenarios considered here, paleo-detectors can probe low-mass WIMPs with masses $m_\chi \lesssim 10\,$GeV even for WIMP-nucleon cross sections many orders of magnitude below current upper bounds. For heavier WIMPs $m_\chi \gtrsim 30\,$GeV the projected sensitivity is better than current limits by a factor of a few to $\sim 100$, depending on the type of interaction and target material. Note that since the dominant background source for such WIMP masses is radioactivity, the sensitivity may be improved substantially with respect to what is presented here if target materials with lower concentrations of uranium than assumed in this work can be obtained. 

We reserve Sec.~\ref{sec:Discussion} for a summarizing discussion.

%*************************
\section{DM Signal}\label{sec:Sig}
%*************************
The differential nuclear recoil rate per unit target mass for elastic scattering of WIMPs with mass $m_\chi$ off nuclei $T$ with mass $m_T$ is
\begin{equation}\label{eq:dRdE1}
 \left( \frac{dR}{dE_R} \right)_T = \frac{2 \rho_\chi}{m_\chi} \int d^3v\, v f({\bf v},t) \frac{d \sigma_T}{dq^2}(q^2,v) \;.
\end{equation}
Here, $E_R$ is the recoil energy of $T$, $\rho_\chi$ the local WIMP density, $f({\bf v},t)$ the WIMP velocity distribution and $d\sigma_T/dq^2$ the differential WIMP-nucleus scattering cross section with the (squared) momentum transfer $q^2 = 2 m_T E_R$. Taking into account canonical spin-independent (SI) and spin-dependent (SD) interactions only, the differential scattering cross section can be written as~\cite{Engel:1991wq,Engel:1992bf,Ressell:1993qm,Bednyakov:2004xq,Bednyakov:2006ux}
\begin{equation}\label{eq:diffxsec}
 \frac{d\sigma_T}{dq^2} (q^2,v) = \frac{1}{v^2} \left[ \frac{\sigma_T^{\rm SI}(0) F^2_{\rm SI}(q^2)}{4 \mu_T^2} + \frac{S_T^{\rm SD}(q^2)}{2J + 1} \right] \Theta(q_{\rm max} - q) , 
\end{equation}
where $\mu_{T}$ is the reduced mass of the WIMP-nucleus system, $\sigma_T^{\rm SI}(0)$ the spin-independent WIMP-nucleus scattering cross section at zero momentum transfer, $F_{\rm SI}$ the nuclear form factor for SI scattering, $S_T^{\rm SD}$ parameterizes the SD WIMP-nucleus scattering cross section, and $J$ is the nuclear spin. The Heaviside $\Theta$-function accounts for the maximal momentum transfer $q_{\rm max} = 2 \mu_T v$ and can be traded for a lower cutoff $v_{\rm min} = \sqrt{m_T E_R / 2 \mu_T^2}$ in the velocity integral. Note that more general WIMP-nucleus scattering operators can lead to velocity dependence different from $d\sigma_T/dq^2 \propto v^{-2}$, see e.g. the non-relativistic effective field theory approach to direct detection~\cite{Fan:2010gt,Fitzpatrick:2012ix}.

With Eq.~\eqref{eq:diffxsec} the differential recoil rate in Eq.~\eqref{eq:dRdE1} can be written as
\begin{equation}\begin{split}\label{eq:dRdE2}
 \left( \frac{dR}{dE_R} \right)_T &= \frac{2 \rho_\chi}{m_\chi} \left[ \frac{\sigma_T^{\rm SI}(0) F^2_{\rm SI}(q^2)}{4 \mu_T^2} + \frac{S_T^{\rm SD}(q^2)}{2J + 1} \right] \\
 & \qquad \times \int\limits_{v_{\rm min}} d^3v\, \frac{f({\bf v},t)}{v} \;.
\end{split}\end{equation}
For this work, we adopt a Maxwell-Boltzmann distribution truncated at the galactic escape velocity $v_{\rm esc}$ and boosted to the Earth's rest frame for the velocity distribution as in the Standard Halo Model~\cite{Drukier:1986tm,Lewin:1995rx,Freese:2012xd}. The remaining velocity integral in Eq.~\eqref{eq:dRdE2} can be calculated analytically, cf. Refs.~\cite{Gondolo:2002np,Catena:2015vpa},
\begin{equation} \begin{split} \label{eq:eta}
 &\quad \int\limits_{v_{\rm min}} d^3v\; \frac{f({\bf v},t)}{v} \\
 &= \frac{1}{N_{\rm esc}}\left\{ \frac{1}{2v_E} \left[ {\rm erf} \left(\frac{v_+}{\sqrt{2}\sigma_v}\right) - {\rm erf}\left(\frac{v_-}{\sqrt{2}\sigma_v} \right) \right] \right. \\
 &\qquad\qquad\quad \left. - \left(\frac{v_+ - v_-}{\sqrt{2\pi} v_E \sigma_v}\right) e^{-\frac{v_{\rm esc}^2}{2\sigma_v^2}} \right\},
\end{split}\end{equation}
with the normalization factor accounting for the truncation of the Maxwell-Boltzmann distribution
\begin{equation}
 N_{\rm esc} = {\rm erf}\left(\frac{v_{\rm esc}}{\sqrt{2}\sigma_v}\right) - \sqrt{\frac{2}{\pi}} \frac{v_{\rm esc}}{\sigma_v} e^{-\frac{v_{\rm esc}^2}{2 \sigma_v^2}} \;,
\end{equation}
and
\begin{equation}
 v_\pm \equiv \min\left(v_{\rm min} \pm v_E, v_{\rm esc} \right) \;.
\end{equation} 
We adopt fiducial values of $\sigma_v = 166\,$km/s for the velocity dispersion and $v_{\rm esc} = 550\,$km/s for the escape velocity. Since the exposure time of paleo-detectors to WIMP-induced nuclear recoils is much larger than the orbital period of the Earth around the Sun, we can neglect the relative motion of the Earth with respect to the Sun. Then, the speed of the detector with respect to the galactic rest frame is given by
\begin{equation}
 v_E = v_\odot \;,
\end{equation}
and we adopt $v_\odot = \sqrt{2} \sigma_v + 13\,{\rm km/s} = 248\,$km/s. 

Note that for exposure times $\mathcal{O}(1)\,$Gyr the relative speed of the Sun with respect to the galactic rest frame may become time dependent. In addition, including the motion of the Earth may have (minor) impact on the sensitivity: the Earth's orbital motion would lead to a broadening of the WIMP induced recoil (and in turn track length) spectra. We leave the investigation of such effects for future work.

%*************************
\subsection{WIMP-Nucleus Interaction Cross Section}
In Eq.~\eqref{eq:diffxsec} we parameterized the differential WIMP-nucleus cross section in terms of the SI and SD cross sections. The zero-momentum WIMP-nucleus SI cross section can be further parameterized as
\begin{equation}
 \sigma_T^{\rm SI}(0) = \frac{4}{\pi}\mu_T^2 \left[ Z_T f_p + \left( A_T-Z_T \right) f_n \right]^2 \;,
\end{equation}
where $A_T$ ($Z_T$) is the number of nucleons (protons) in the target nucleus and $f_p$ ($f_n$) is the effective WIMP-proton (WIMP-neutron) coupling. Many WIMP models exhibit nearly isospin-conserving WIMP-nucleon couplings $f_p \approx f_n$, leading to the typical $\sigma_T^{\rm SI}(0) \propto A_T^2$ enhancement in the WIMP-nucleus scattering cross section. The couplings $f_p$ ($f_n$) are related to the WIMP-nucleon cross sections $\sigma_{p,n}^{\rm SI}$ (the quantity in which experimental results and projections are often quoted) by $\sigma_{p,n}^{\rm SI} = 4 \mu_{p,n}^2 f_{p,n}^2/\pi$ with $\mu_p$ ($\mu_n$) the WIMP-proton (WIMP-neutron) reduced mass. 

We use the Helm nuclear form factor~\cite{Helm:1956zz,Lewin:1995rx,Duda:2006uk} for SI scattering
\begin{equation}
 F_{\rm SI}(q^2) = 3 \frac{\sin(q r_n) - q r_n \cos(q r_n)}{(q r_n)^3} e^{- (q s)^2/2} \;,
\end{equation}
with the effective nuclear radius $r_n^2 \approx c^2 + \frac{7}{3} \pi^2 a^2 - 5 s^2$ where $a \approx 0.52\,$fm, $c \approx (1.23 A_T^{1/3} - 0.6)\,$fm, and $s \approx 0.9\,$fm. Note that more refined calculations of the form factors are available, although only for a few isotopes, see e.g. Refs.~\cite{Vietze:2014vsa,Gazda:2016mrp,Korber:2017ery,Hoferichter:2018acd}.

For SD interactions it is useful to decompose the interactions into an isoscalar ($S_{00}^T$), an isovector ($S_{11}^T$), and an interference term ($S_{01}^T$)~\cite{Bednyakov:2004xq,Bednyakov:2006ux}
\begin{equation} \label{eq:SSD_qd}
 S_T^{\rm SD}(q^2) = a_0^2 S^T_{00}(q^2) + a_1^2 S^T_{11}(q^2) + a_0 a_1 S^T_{01}(q^2) \;,
\end{equation}
with the effective isoscalar and isovector couplings
\begin{equation}
 a_0 \equiv a_p + a_n \;, \qquad a_1 = a_p - a_n \;.
\end{equation}
The effective SD WIMP-nucleon couplings are related to the WIMP-nucleon scattering cross section as $\sigma_{p,n}^{\rm SD}(0) = 12 \mu_{p,n}^2 a_{p,n}^2/\pi$.

Where available\footnote{For the isotopes \{ $^{19}$F, $^{23}$Na, $^{27}$Al, $^{29}$Si, $^{39}$K, $^{73}$Ge, $^{93}$Ni, $^{123}$Te, $^{125}$Te, $^{127}$I, $^{129}$Xe, $^{131}$Xe, $^{207}$Pb\}} we use the structure functions $S^T_{ij}$ as tabulated in~\cite{Bednyakov:2006ux}. For target nuclei where such structure functions do not exist we use the simplified parameterization of the SD WIMP-nucleon cross section
\begin{equation} \label{eq:SSD_q0}
 S_T^{\rm SD}(q^2) = \frac{\left(2J+1\right)}{\pi} \frac{\left(J+1\right)}{J} F^2_{\rm SD}(q^2) \left[ a_p \langle{\bf S}_p^T\rangle + a_n \langle{\bf S}_n^T\rangle \right]^2 ,
\end{equation}
where $\langle{\bf S}_p^T\rangle$ and $\langle{\bf S}_n^T\rangle$ are the proton and neutron spin, respectively, averaged over the nucleus. We use values for the $\langle{\bf S}_i^T\rangle$ as tabulated in Ref.~\cite{Bednyakov:2004xq}. Note that there are considerable differences for these values when calculated in different nuclear models; we follow the recommendations given in Ref.~\cite{Bednyakov:2004xq} for which set of values to use for each isotope. 

For the nuclear structure function, we use the form given in Ref.~\cite{Lewin:1995rx}
\begin{equation}
 F^2_{\rm SD}(qr_n) = \begin{cases} 0.047 &,{\rm ~for}\;2.55\leq qr_n\leq 4.5 \;,\\
 \sin^2(qr_n)/(q r_n)^2 &,{\rm ~else}\;, \end{cases}
\end{equation}
with the effective nuclear radius $r_n = 1.0 A_T^{1/3}\,$fm. As in the SI case, more refined calculations of the form factors are available for some target isotopes, see e.g. Refs.~\cite{Klos:2013rwa,Stroberg:2016ung,Gazda:2016mrp,Sahu:2017czz}.

%*************************
\subsection{Track-Length Estimate} \label{sec:Sig_TrackLength}

From the differential recoil rate for given target nuclei we compute the associated spectrum of damage track lengths. In obtaining our results, track lengths are obtained using numerical calculations described below. To illustrate the relevant characteristics of the track length calculation, we begin by presenting a semi-analytic approximation here. The stopping power for a recoiling nucleus $T$ incident on an amorphous target $V$ with atomic number density $n_V$ can be estimated by~\cite{Wilson:1977jzf} 
\begin{equation}\label{eq:dEdx}
 \left( \frac{dE}{dx}\right)_{TV} = n_V \frac{\pi a_{TV}^2 \gamma_{TV}}{C_{TV}} S(\epsilon_{TV})\;,
\end{equation}
with the reduced energy
\begin{equation}
 \epsilon_{TV} = \frac{\mu_{(TV)}}{m_T} \frac{a_{TV} E}{\alpha Z_T Z_V} \;.
\end{equation}
$Z_{T/V}$ denotes the number of protons in $T/V$, $C_{TV} = \epsilon_{TV}/E$, $\gamma_{TV} = 4 \mu_{TV}$ where $\mu_{(TV)}$ is the reduced mass of the $T$--$V$ system, and $\alpha$ is the fine-structure constant. The reduced stopping power $S$ can be derived from a screened interatomic Coulomb potential with screening length
\begin{equation}
 a_{TV} = 0.8853 a_0 / \left( \sqrt{Z_T} + \sqrt{Z_V} \right)^{2/3} \;,
\end{equation}
where $a_0$ is the Bohr radius. For an average of various interatomic potentials, the reduced stopping power can be parameterized as~\cite{Wilson:1977jzf}
\begin{equation}\label{eq:redScreening}
 S(\epsilon) \approx \frac{1}{2} \frac{\ln(1+\epsilon)}{\left( \epsilon + A \epsilon^B \right)} + k \sqrt{\epsilon} \;,
\end{equation}
where the parameters $A=0.14120$, $B=0.42059$ and $k=0.15$ yield a reasonable fit to data for a wide variety of $T/V$ combinations. Note that the first term arises from nuclear stopping and dominates when $\epsilon \ll 1$, while the second term arises from stopping due to the electrons associated with the target nuclei.

For a composite material, the stopping power is obtained by summing over the contributions from different constituents $V$,
\begin{equation} \label{eq:CompStop}
 \frac{dE}{dx_T} = \sum_V \left( \frac{dE}{dx}\right)_{TV},
\end{equation}
and the track length for a recoiling nucleus with energy $E_R$ is
\begin{equation} \label{eq:TrackLength}
 x_T(E_R) = \int_0^{E_R} dE \left( \frac{dE}{dx_T}(E) \right)^{-1} \;.
\end{equation}
The track length spectrum for WIMP-induced nuclear recoils within a target mineral is then given by a sum over constituent nuclei
\begin{equation} \label{eq:CompStop}
 \frac{dR}{dx} = \sum_T \xi_T \frac{dE_R}{dx_T} \left(\frac{dR}{dE_R}\right)_T ,
\end{equation}
where $\xi_T$ is the mass fraction of the nuclei.

\begin{figure}
 \includegraphics[width=\linewidth,trim={-0.64cm, 1.2cm, 0cm, 0cm},clip]{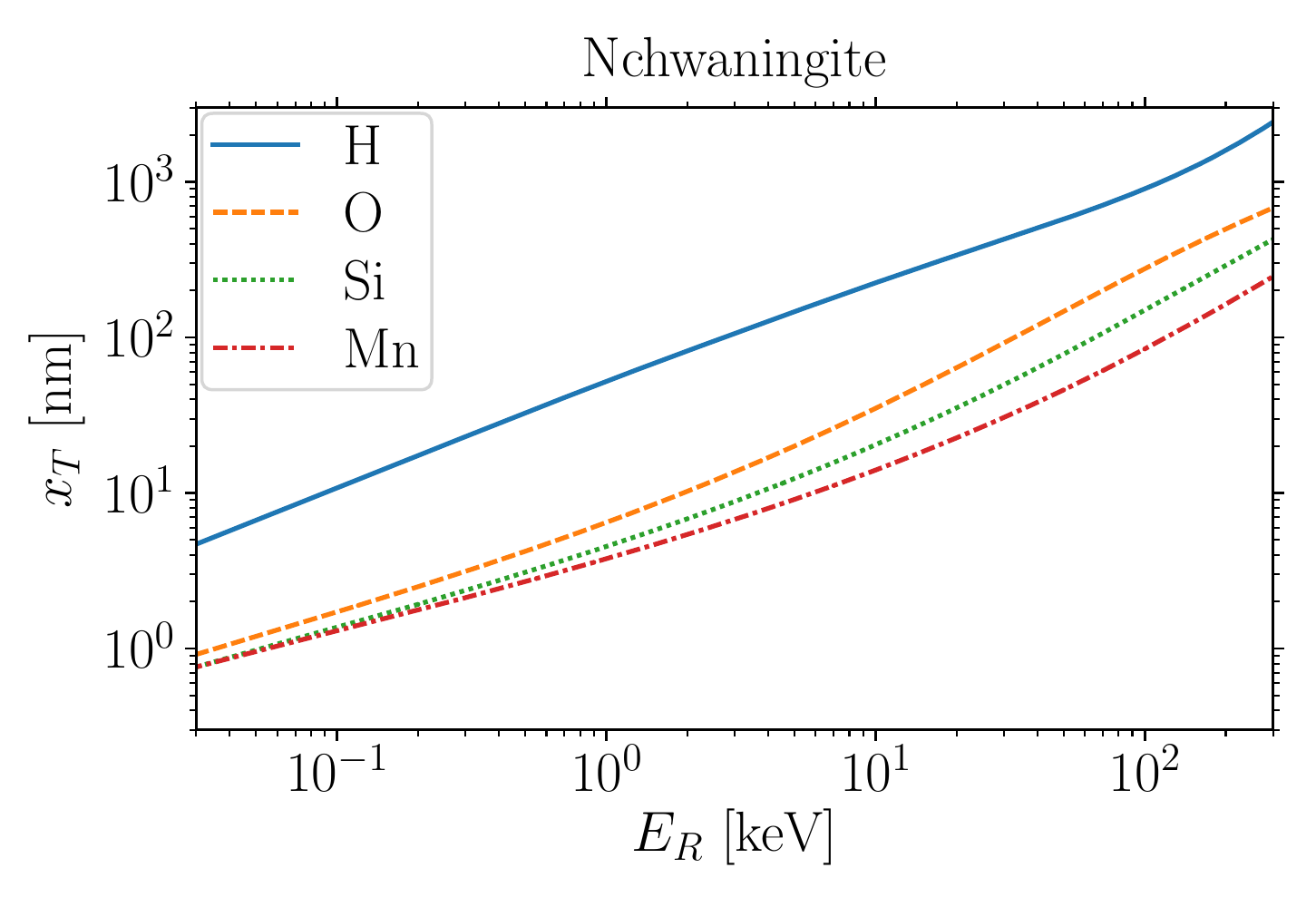}

 \includegraphics[width=\linewidth,trim={0cm, 0cm, 0cm, 1cm},clip]{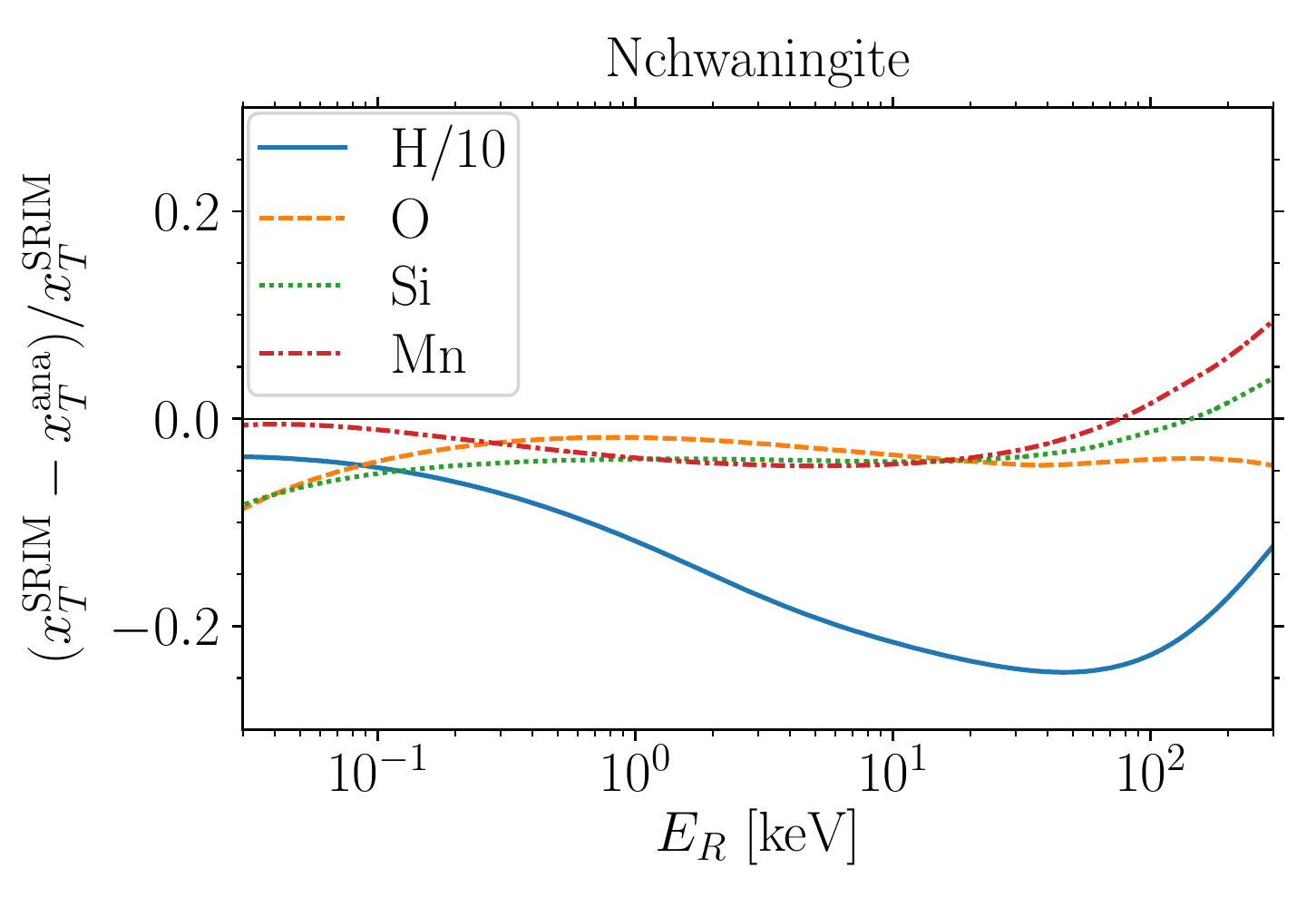}
 \caption{{\it Top:} Track length $x_T$ as a function of nuclear recoil energy $E_R$ for the different target nuclei in Nchwaningite [\ccNch] calculated with \texttt{SRIM}. {\it Bottom:} Relative difference of $x_T^{\rm ana}(E_R)$ from the semi-analytic calculation and $x_T^{\rm SRIM}$ calculated with \texttt{SRIM}. Note that for hydrogen the relative difference is rescaled by a factor $1/10$. For recoil energies outside of $0.1\,{\rm keV} \lesssim E_R \lesssim 100\,$keV  the differences between the semi-analytic calculation and the \texttt{SRIM} results become sizable. Further, the two computations differ widely for low-$Z$ nuclei such as hydrogen. As discussed further in the text, this stems from the semi-analytic calculation being matched to experimental data only for larger-$Z$ nuclei and recoil energies $0.1\,{\rm keV} \lesssim E_R \lesssim 100\,$keV, while the \texttt{SRIM} results describe experimental data well for larger/smaller recoil energies and a larger variety of nuclei.}
 \label{fig:Nch_Range}
\end{figure}

\begin{figure*}
 \includegraphics[width=0.49\linewidth]{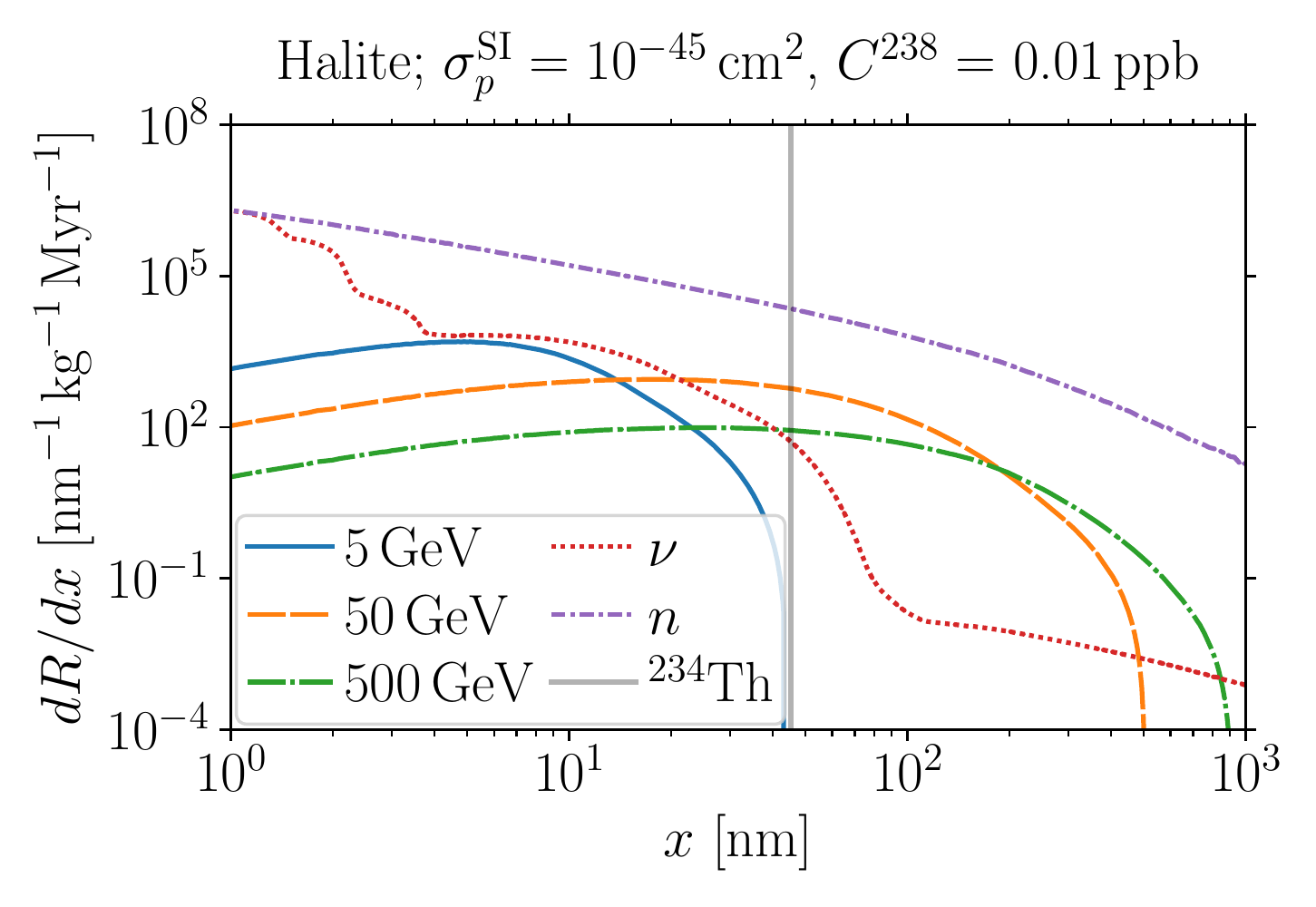}
 \includegraphics[width=0.49\linewidth]{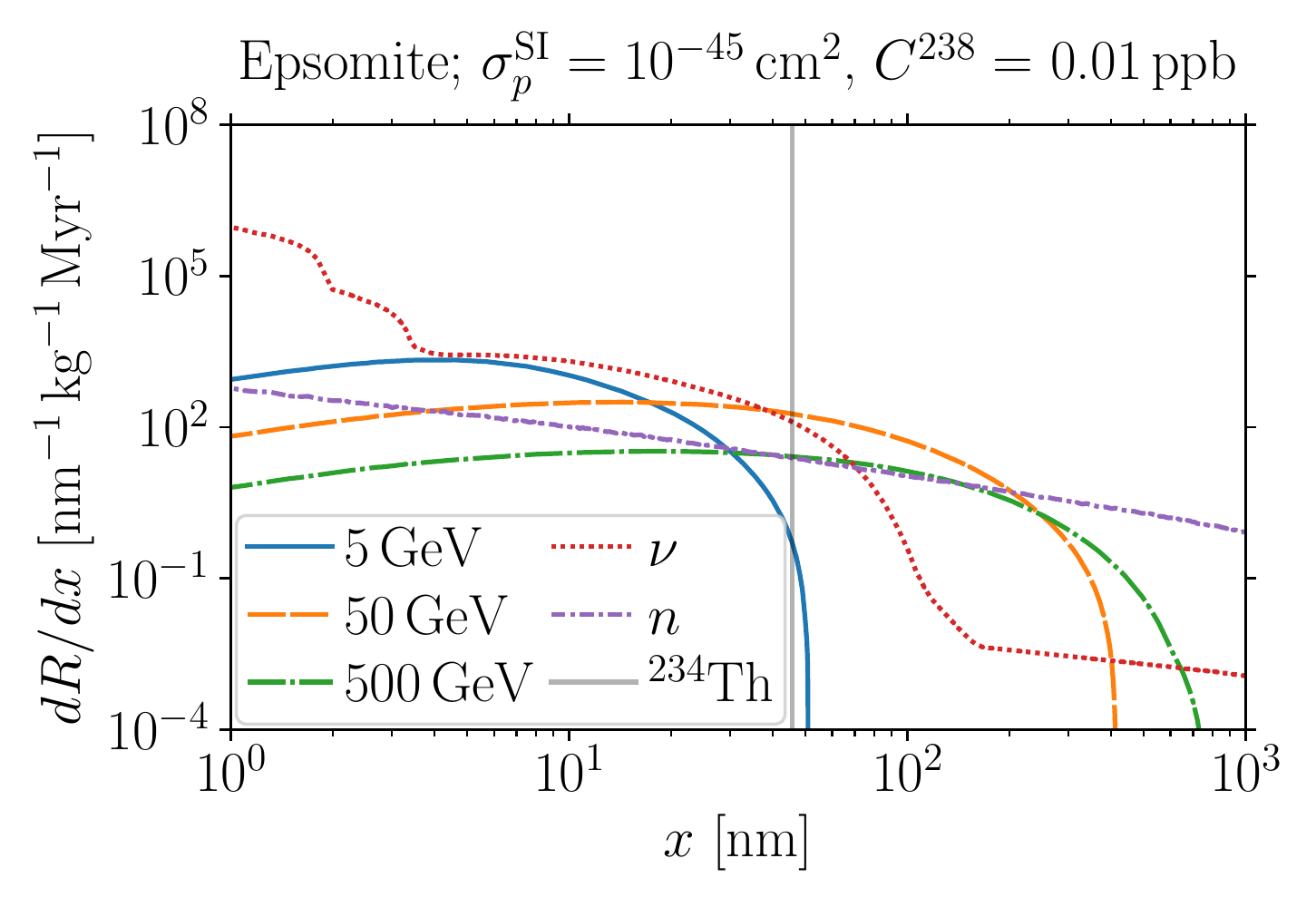}

 \includegraphics[width=0.49\linewidth]{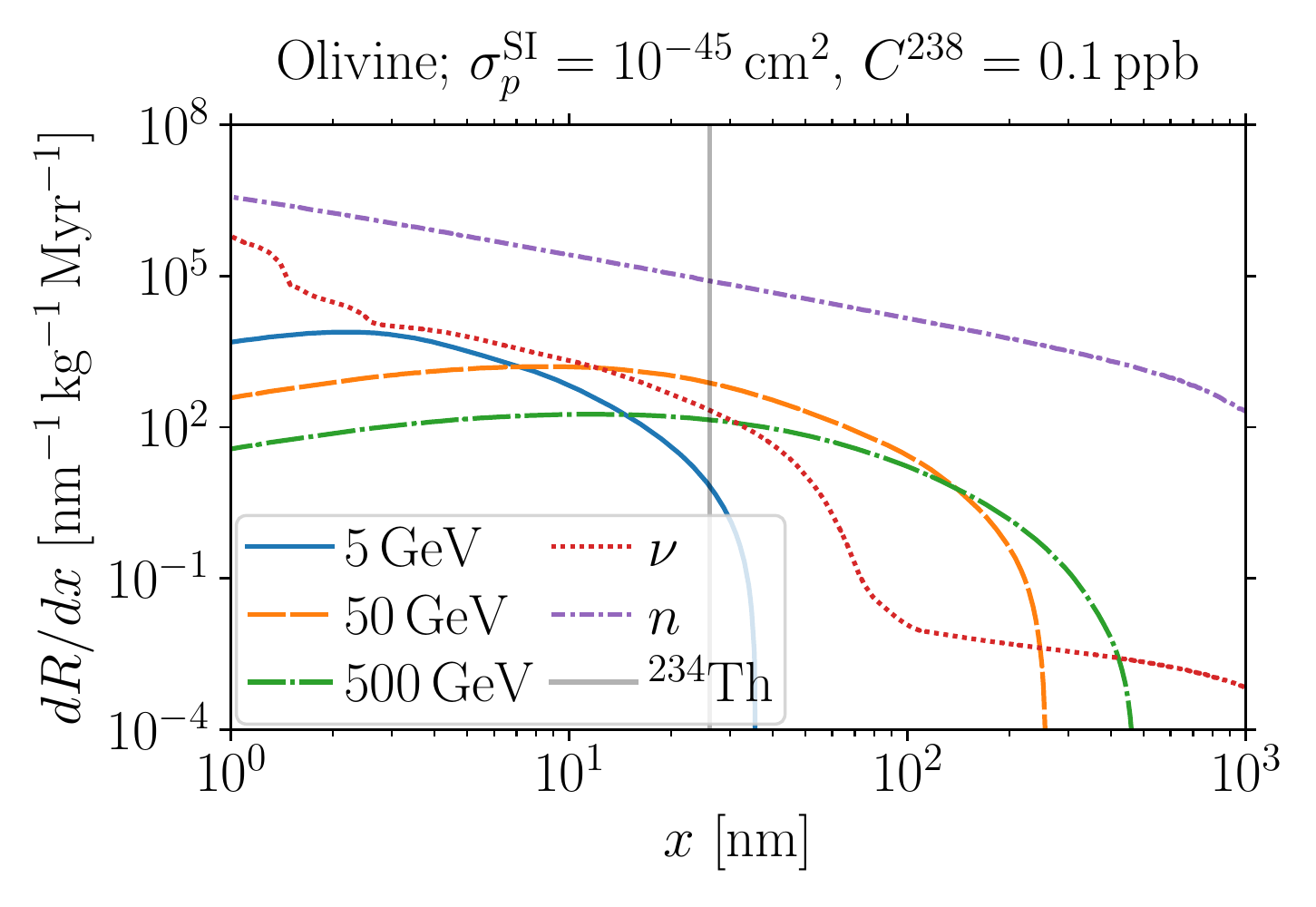}
 \includegraphics[width=0.49\linewidth]{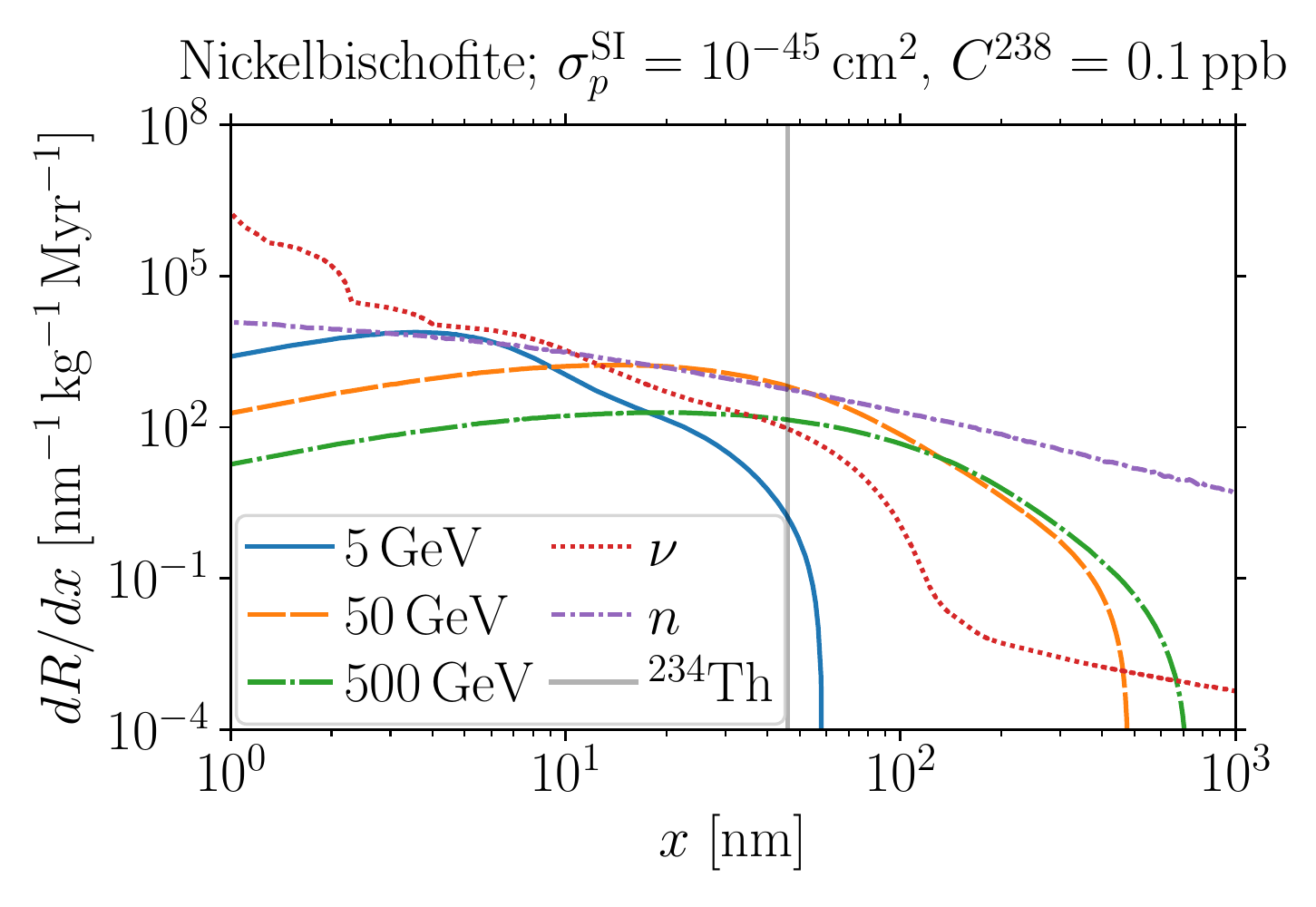}
 \caption{Track length spectra from recoiling nuclei in Halite (\ccHal; top left), Epsomite [\ccEps; top right], Olivine [\ccOli; bottom left], and Nickelbischofite [\ccNic; bottom right] induced by the neutrino ($\nu$) and neutron ($n$) background and WIMPs with $m_\chi = \{5,\;50,\;500\}\,$GeV, assuming $\sigma_p^{\rm SI} = 10^{-45}\,{\rm cm}^2$. The vertical gray line indicates the track length of $72\,$keV $^{234}$Th nuclei from $(^{238}{\rm U} \to {^{234}{\rm Th}} + \alpha)$ decays. Halite and Epsomite are MEs for which we assume a $^{238}$U concentration of $C^{238}=0.01\,$ppb in weight while for the UBRs Olivine and Nickelbischofite we assume $C^{238}=0.1\,$ppb. See Sec.~\ref{sec:Bkg} for a discussion of the background sources.}
 \label{fig:Spectra}
\end{figure*}

While illustrative of the electronic and nuclear stopping of recoiling nuclei in amorphous targets, the parametrization of the stopping power used in the semi-analytic approximation outlined above, in particular the functional form of the reduced stopping power Eq.~\eqref{eq:redScreening} and the values of the numerical coefficients appearing therein, are only matched to previous experimental data for particular nuclei and energy ranges. In obtaining our results we instead use the stopping power obtained from the \texttt{SRIM} code~\cite{Ziegler:1985,Ziegler:2010} and calculate the associated track lengths using Eq.~\eqref{eq:TrackLength}. Although the \texttt{SRIM} code improves the semi-analytic treatment by taking into account data from a more complete collection of nuclei at much wider ranges of energy, these results strictly still hold for amorphous targets only, i.e. neglect effects from the crystalline nature of our targets such as channeling~\cite{Bozorgnia2010a,Bozorgnia2010b}.

We show the track length as a function of recoil energy for the different target nuclei in Nchwaningite [\ccNch] in the top panel of Fig.~\ref{fig:Nch_Range}. Note that lighter nuclei give rise to significantly longer tracks than heavier nuclei for the same recoil energy because the stopping power increases with the charge of the nucleus, cf. Eqs.~\eqref{eq:dEdx}-\eqref{eq:redScreening}. Such figures look similar for other target nuclei, with quantitative differences mainly arising from varying molecular number densities between different target minerals. In general, lower molecular number densities yield longer tracks for the same recoiling nucleus and recoil energy, cf. Eq.~\eqref{eq:dEdx}. 

The top panel of Fig.~\ref{fig:Nch_Range} also indicates the smallest nuclear recoils which could be detected with paleo-detectors: If tracks as short as $\mathcal{O}(1)\,$nm can be reconstructed, paleo-detectors are sensitive to nuclear recoils as small as $\mathcal{O}(100)\,$eV. Read-out methods with somewhat worse spatial resolution, say $\mathcal{O}(10)\,$nm, would correspond to a recoil energy threshold of order keV.

In the bottom panel of Fig.~\ref{fig:Nch_Range} we show the relative difference of the track length spectra calculated with the semi-analytic approximation outlined above and with the \texttt{SRIM} code. In the most relevant recoil energy range $0.1\,{\rm keV} \lesssim E_R \lesssim 100\,$keV both results differ by less than 10\,\% for nuclei other than hydrogen. We find larger differences outside of this energy range as well as for low-$Z$ nuclei such as hydrogen. Note that the stopping power in the semi-analytic calculation is not fit to data outside of this energy range or for low-$Z$ recoils, while the results of the \texttt{SRIM} code still describe experimental data well in this region.

In Fig.~\ref{fig:Spectra} we show the track length spectra induced by WIMPs in different target materials together with background spectra discussed in the following section. We show spectra induced in two marine evaporites (MEs), Halite (\ccHal) and Epsomite [\ccEps] and two ultra-basic rocks (UBRs), Olivine [\ccOli] and Nickelbischofite [\ccNic].\footnote{Here and in the remainder of this paper, for brevity, we refer to minerals in marine evaporite deposits as ``marine evaporites'' (MEs) and to minerals in ultra-basic rocks as ``ultra-basic rocks'' (UBRs).} For both MEs and UBRs this selection contains one mineral with and one without hydrogen. Note that we do not include tracks from hydrogen in the track length spectra shown in Fig.~\ref{fig:Spectra}, cf. the discussion in Sec.~\ref{sec:ReadOut}. Also, heavier WIMPs give rise to harder track length spectra than lighter WIMPs because heavier WIMPs induce more energetic nuclear recoils.

Note that the track length discussed above is strictly the range of nuclei in the material. In the following, we will assume that this range coincides with the measured track length modulo errors induced by the finite spatial resolution of the particular read out method. In principle, this assumption can be violated either if permanent damage only arises in part of a nucleus' range, or if the nucleus loses energy via hard scatterings such that the shape of the track deviates significantly from a straight line. Both cases could lead to reconstructed tracks appearing much shorter than the range calculated above. Measured track lengths shorter than the calculated range could lead to diminished sensitivity of paleo-detectors, although such effects can be partly mitigated by improved spatial resolution or larger exposures.  

In any case, our studies using \texttt{SRIM} indicate that corrections to the tracks lengths due to inconsistent appearance of persistent damage or to significant scattering of nuclei are small. In particular, they play a role only at the end of the track when the nucleus has already lost most of its energy and traversed most of its range. The effect of such corrections is a question which currently cannot be answered quantitatively; detailed experimental studies are required for each target material, which we leave for future work. To the best of our knowledge, reliable estimates exist only for the particular case of reconstructing tracks in muscovite mica after cleaving and chemical etching~\cite{Collar:1994mj}.

%*************************
\section{Read-out Methods}\label{sec:ReadOut}
%*************************

In order to be suitable for paleo-detectors, target minerals must satisfy a few basic criteria. Foremost, nuclei with energies of order $E_R = 0.1-100\,$keV should give rise to damage tracks in the material, and such tracks must persist over sufficiently long time scales. Materials with such properties are commonly referred to as Solid State Track Detectors (SSTDs)~\cite{Fleischer:1964,Fleischer383,Fleischer:1965yv,GUO2012233}. 

The exact mechanism for track formation by ions in solids is not fully understood; popular models include the thermal spike model~\cite{Seitz:1949} and the ion explosion model~\cite{Fleischer:1965}. Which ions leave permanent tracks in which materials remains a question that can only be answered semi-empirically. In general, materials must be insulators (or poor semiconductors) with electrical resistivity larger than $\sim 2000\,\Omega\,$cm to record tracks, see Ref.~\cite{GUO2012233} for a more detailed discussion of track formation criteria. Much of research using SSTDs has employed optical microscopy to read out tracks after chemical etching. Tracks from ions with charge $Z \gtrsim 10$ and energies larger than a few keV are easily etchable in virtually all insulating materials studied. For tracks from lighter nuclei, in particular $\alpha$-particles ($^4$He$^{2+}$ ions), the situation is less clear. Proton ($^1$H$^+$) tracks have been demonstrated to be etchable only in plastics~\cite{GUO2012233}. However, tracks from $\alpha$-particles with $\sim 5\,$MeV energies have been measured in (Al$_2$O$_3$:C,Mg) crystals without etching using confocal laser scanning and structured illumination microscopy~\cite{BARTZ2013273,Kouwenberg:2018}. In this work we remain agnostic about the question of low-$Z$ particles leaving tracks in the targets, but will instead consider two scenarios: {\it with low-$Z$ tracks}, where we assume that all ions leave tracks, and {\it without low-$Z$ tracks} where we assume that only ions with $Z \geq 3$ leave observable tracks.\footnote{For paleo-detectors the most relevant distinction is if hydrogen nuclei leave tracks or not. Recovering tracks from $\alpha$-particles or not has almost no effect on the projected sensitivity of the materials considered here.} Which scenario applies will have to be determined experimentally for each combination of target material and read-out method.

Besides recording tracks in the first place, materials also have to preserve tracks over geological timescales of $0.1-1\,$Gyr in order to be suitable for use as paleo-detectors. The fading, or {\it annealing}, of tracks is thought to be due to the diffusion of atoms from the vicinity of the damaged region into the tracks. The annealing time scale is exponentially suppressed with the temperature $T$ of the material~\cite{Fleischer:1965yv,GUO2012233}
\begin{equation}
 t_{\rm ann} \propto e^{E_{\rm ann}/kT} \;,
\end{equation}
where $E_{\rm ann}$ is the so-called activation energy for annealing and $k$ the Boltzmann constant. Typically, $E_{\rm ann} \sim k T_m$, where $T_m$ is the melting temperature. For most minerals the annealing time at room temperature is much larger than 1\,Gyr. For typical low melting-point crystals, e.g. Calcite, annealing times are $t \gtrsim 10^9\,$yr at room temperature. Because of their high melting temperatures, refractory materials are of particular interest because they typically display large $E_{\rm ann}$ and in turn annealing times, e.g. for Diopside (CaMgSi$_2$O$_6$) the annealing time has been estimated to be $t \sim 10^{59}\,$yr at room temperature~\cite{Fleischer:1965yv}. Note that reported temperatures at the bottom of ultra-deep boreholes vary from $\sim 100\,^\circ{\rm C}$ to $\sim 300\,^\circ{\rm C}$, depending on the depth and local geology. Thus, depending on the temperatures at the sites from where target materials for paleo-detectors will be obtained, track annealing may play a role for low melting-point materials. In refractory materials, annealing times are larger than 1\,Gyr even at temperatures of a few hundred degrees Celsius.

Damage tracks which persist over geological time scales have been studied extensively for {\it fission track dating}, typically making use of tracks caused by the $Z \sim 50$ fragments from spontaneous fission of $^{238}$U. A commonly used mineral for fission track dating is muscovite mica, which can be cleaved into thin slices. Fission tracks are then read out with e.g. Electron Microscopy (EM) or after chemical etching with optical microscopes. Note that the $\sim\,$nm resolution of EM allows the imaging of fission tracks without prior enlargement by chemical etching. However, chemical etching is often used for EM read-out in order to stabilize tracks against thermal annealing caused by the incident electron beam.~\cite{Fleischer:1965yv}

Previous searches for DM-induced nuclear recoil tracks used techniques similar to those of fission track dating, employing an Atomic Force Microscope (AFM) to scan the surfaces of cleaved and etched samples of muscovite mica~\cite{SnowdenIfft:1995ke}. These past experiments were limited by both the small throughput allowed by the extensive time required for sample preparation and the manifestly two-dimensional nature of the ion track reconstruction. 

We propose to read out samples with recently developed Helium Ion beam Microscopy (HIM)~\cite{HILL201265}, which has spatial resolution similar to EM. However, it causes less sample damage than EM~\cite{VANGASTEL20122104} and is capable of sub-surface imaging to depths of $\mathcal{O}(100)\,$nm. Further progress with respect to the two-dimensional readout with EM or AFM can be made by using a Focused Ion Beam (FIB) of either neon or gallium ions for sample preparation (for example, see~\cite{Lombardo:pp5019}). The FIB can both be used for initial sample preparation and for realizing three dimensional reconstruction of samples by sequentially imaging a surface layer with HIM and then sputtering away the read-out layer using the FIB~\cite{bioHIB}. Much faster ablation of the read-out layer is achievable by using pulsed lasers in addition to the FIB~\cite{ECHLIN20151,PFEIFENBERGER2017109,Randolph:2018}. We estimate that combining HIM with pulsed laser ablation\footnote{In practice, one would use a combination of pulsed-laser and ion-beam ablation. Most of the previously read-out layer can be ablated with pulsed lasers to maximize throughput. However, laser ablation potentially also causes substantial local thermal annealing of damage tracks. Thus, one would use ion-beams of descending $Z$-nuclei to ablate the remaining portion of the previously read-out layer, minimizing thermal annealing in the layer to be read-out next.}, target volumes of a few ${\rm mm}^3$, corresponding to masses of $\mathcal{O}(10)\,$mg, should be possible. Although the target masses which can be read out with HIM are particularly small, we see from Fig.~\ref{fig:Spectra} that, for $m_\chi \lesssim 10\,$GeV, WIMP-induced recoils yield $\lesssim 10\,$nm long nuclear recoil tracks at higher rates when compared to heavier WIMPs. Thus, the $\sim\,$nm spatial resolution possible with HIM read-out is ultimately more relevant for sensitivity to low mass WIMPs since, as with conventional DD experiments, paleo-detectors become limited by the threshold of the detector rather than its exposure. 

Alternatively, for higher mass WIMPs with significantly longer induced nuclear recoil tracks, Small Angle X-ray scattering (SAXs) at synchrotron facilities should be capable of fully three dimensional imaging of bulk samples with minimal sample preparation~\cite{SAXS3d}. Ion tracks have been revealed without etching in crystalline materials using SAXs, although only after imaging the sample along the direction of the recorded track~\cite{RODRIGUEZ2014150}. Also, SAXs tomography has achieved $\sim 15\,$nm three-dimensional spatial resolution~\cite{SAXSres} but not for resolving damage from ion tracks, which cause only small variations in electron density of the target material. Thus, while we are proposing a particularly challenging application of SAXs and the available beam time can be limited, we estimate a total target volume of a few tens of cm$^3$, corresponding to $\mathcal{O}(100)\,$g, can be imaged at synchrotron facilities. The $\sim 10^4$ increase in exposure relative to HIM allows for SAXs to probe much lower WIMP-nucleon scattering cross sections for WIMPs with $m_\chi \gtrsim 10\,$GeV despite the loss of spatial resolution. 

Even larger target volumes can be handled by Confocal Laser Scanning Microscopy (CLSM) or Structured Illumination Microscopy (SIM), although at the cost of worse spatial resolution of the order of $100\,$nm. Reliable read out of $\alpha$-particle tracks with sufficient spatial resolution to allow for energy spectroscopy of the incident $\alpha$-flux has been demonstrated with CLSM and SIM without prior chemical etching in particular target materials (Al$_2$O$_3$:C,Mg)~\cite{BARTZ2013273,Kouwenberg:2018}. Somewhat better spatial resolution may be achievable with Ultra-Violet Microscopy (UVM).

Read-out methods which can process comparatively large target volumes such as CLSM, SIM, or UVM are particularly useful to pre-screen target materials for paleo-detectors and identify sub-volumes which are low in radioactively induced backgrounds. Such sub-volumes could then be interrogated for traces of WIMP interactions with higher resolution read-out methods such as SAXs or HIM.

%*************************
\section{Background Rejection}\label{sec:Bkg}
%*************************
In order to be sensitive to DM signals, a number of background sources must be mitigated or controlled to a sufficient level in paleo-detectors. Most of the background sources are the same as in the conventional direct detection approach. However, the relative importance of the respective background sources is different in paleo-detectors and conventional direct detection experiments due to three reasons:
\begin{enumerate}[1)]
 \item Paleo-detectors use long exposure times \mbox{[$\lesssim \mathcal{O}(10^9)\,$yr]} and small target masses \mbox{[$\lesssim \mathcal{O}(1)\,$kg]} while conventional direct detection experiments use short exposure times \mbox{[$\lesssim \mathcal{O}(10)\,$yr]} and large target masses \mbox{[$\lesssim \mathcal{O}(10^4)\,$kg]}.
 \item The experimental observable in paleo-detectors is $1-500\,$nm long damage tracks from recoiling nuclei, while conventional direct detection experiments observe the ionization charge, the scintillation light, or the heat (phonons) produced by the recoiling nucleus.
 \item Paleo-detectors measure events integrated over the exposure time while conventional experiments have precise timing information.
\end{enumerate}
In the remainder of this section we discuss the dominant sources of backgrounds and how to mitigate or control them. Note that natural crystal imperfections are either single-site or span across the entire (mono-)crystalline volume and are thus easily rejected. All relevant backgrounds are damage tracks from charged particles being stopped in the material.

%*************************
\subsection{Cosmogenic}
Cosmic rays can scatter off target nuclei, leading to recoils similar to WIMP-induced nuclear recoils. Similar to conventional direct detection experiments, such background can be mitigated by using target samples obtained from far below the Earth's surface. The dominant cosmogenic background will then be due to muons interacting with nuclei in the vicinity of the target volume, giving rise to fast neutrons, which in turn scatter off target nuclei, inducing nuclear recoils. 

In conventional direct detection experiments, such neutron-induced nuclear recoils can be mitigated further by rejecting coincident events: fast neutrons have mean free paths of $\mathcal{O}(1)\,$cm in typical detector materials and usually scatter multiple times in the detector volume. Note that because the neutron's mass is much smaller than the mass of typical target nuclei, neutrons lose only a small fraction of their energy in a single scattering event. Due to the lack of timing information, such suppression strategies cannot be employed in paleo-detectors.

However, conventional detectors must be operated in (large) underground laboratories. The deepest such laboratories currently available ({\it Snolab}, Canada and {\it CJPL}, China) have an overburden of $\lesssim 2\,$km of rock corresponding to $\lesssim 7\,$km.w.e. (km water equivalent). At such depths the neutron flux is $\mathcal{O}(10^{-2})\,{\rm m}^{-2}\,{\rm yr}^{-1} = \mathcal{O}(10^3)\,{\rm cm}^{-2}\,{\rm Gyr}^{-1}$~\cite{Mei:2005gm}. While a neutron flux of this magnitude is acceptable for a conventional direct detection experiment it would pose a severe problem for paleo-detectors due to the large exposure and the lack of timing information. However, the neutron flux is exponentially suppressed with the overburden. Target materials for paleo-detectors can be obtained from depths much larger than 2\,km, e.g. from ultra-deep boreholes as used for geological R\&D and oil exploration. The neutron flux at depths of \{5\,,\;7.5\,,\;10\}\,km rock overburden is of order \{$1\,,\;10^{-4}\,,\;10^{-8}$\}\,cm$^{-2}$\,Gyr$^{-1}$, making cosmogenic background negligible for materials obtained from depths larger than $\gtrsim 5\,$km. Note that both MEs (for example, see~\cite{Blattlereaar2687}) and UBRs (for example, see~\cite{Hirschmann1997}) with ages $\gtrsim 500\,$Myr have been found in cores from ultra-deep boreholes.

Due to the comparatively small size of the target sample, near-surface storage of the minerals between when they are obtained from deep in the Earth and read out does not lead to problematic levels of irradiation with cosmic rays. For example, the induced neutron flux in a $\sim 50\,$m deep storage facility is smaller than $\sim 0.2\,$cm$^{-2}$\,yr$^{-1}$.

%*************************
\subsection{Radioactive Decays}

\begin{table}
 \begin{tabular}{C{2cm} C{3cm} C{2.5cm}}
 \hline\hline
 Nucleus & Decay Mode & $T_{1/2}$ \\
 \hline
 \multirow{2}{*}{$^{238}$U} & $\alpha$ & $4.468 \times 10^9$\,yr \\
 & SF & $8.2 \times 10^{15}$\,yr \\
 $^{234}$Th & $\beta^-$ & 24.10\,d \\
 \multirow{2}{*}{$^{234{\rm m}}$Pa} & $\beta^-$ (99.84\,\%) & \multirow{2}{*}{1.159\,min} \\
 & IT (0.16\,\%) & \\
 $^{234}$Pa & $\beta^-$ & 6.70\,d \\
 $^{234}$U & $\alpha$ & $2.455 \times 10^5\,$yr \\
 $^{230}$Th & $\alpha$ & $7.54 \times 10^4\,$yr\\
 $^{226}$Ra & $\alpha$ & $1600\,$yr \\
 $^{222}$Rn & $\alpha$ & $3.8325\,$d \\
 \hline\hline
 \end{tabular}
 \caption{Half-lives $T_{1/2}$ of selected nuclei in the $^{238}$U decay chain~\cite{Browne:2015yci,Browne:2006zz,Browne:2012cxp,AKOVALI:1996ehn,Singh:2011yau}. For $^{238}$U we denote both the half-lives corresponding to $\alpha$-decays and spontaneous fission (SF). For $^{234{\rm m}}$Pa we denote in addition the branching ratio for $\beta^-$-decays and isomeric transitions (IT).}
 \label{tab:U_halflife}
\end{table}

Any target sample used for paleo-detectors will be contaminated by traces of radioactive materials. The mitigation of the corresponding background is crucial for the success of the proposed search. Similar to conventional direct detection experiments, it is crucial to select materials with the lowest possible concentrations of radioactive materials. We discuss typical contaminations of target materials further in Sec.~\ref{sec:MinOpt}. The most relevant contaminant is $^{238}$U. As a benchmark value, we will assume $^{238}$U concentrations of $C^{238} = 0.01\,$ppb (part per billion) in weight in the following discussion.

The $^{238}$U decay chain is
\begin{equation}\begin{split} \label{eq:Uchain}
 &{ ^{238}{\rm U} } \stackrel{\alpha}{\longrightarrow} { ^{234}{\rm Th} } \stackrel{\beta^-}{\longrightarrow} { ^{234{\rm m}}{\rm Pa} } \stackrel{\beta^-}{\longrightarrow} { ^{234}{\rm U}} \stackrel{\alpha}{\longrightarrow} {^{230}{\rm Th}} \\
 & \quad \stackrel{\alpha}{\longrightarrow} {^{226}{\rm Ra}} \stackrel{\alpha}{\longrightarrow} {^{222}{\rm Rn}} \stackrel{\alpha}{\longrightarrow} \ldots \longrightarrow {^{206}{\rm Pb}}\;. 
\end{split}\end{equation}
We quote the half-lives for the most relevant decays in Tab.~\ref{tab:U_halflife}. 

In paleo-detectors, there are qualitative differences between $\alpha$-decays and $\beta/\gamma$-decays: $\alpha$-decays are ($N \to N' + \alpha$) 2-body decays, giving rise to a mono-energetic nuclear recoil and an $\alpha$-particle. For the $\alpha$-decays of heavy nuclei in the uranium and thorium decay chains, the energies of the $\alpha$-particles are a few MeV, and the induced nuclear recoils have energies of $10-100$\,keV. Both the recoiling nucleus and the $\alpha$-particle lose their energy mostly via ionization and elastic scattering off other nuclei, and may give rise to a damage track in the target material. $\beta$ ($\gamma$) decays are 3-body (2-body) decays where the nucleus emits an electron and a neutrino (a photon). Since the mass of the electrons/neutrinos/photons is negligible with respect to the mass of the nucleus, the light decay products carry most of the excess energy of the decay. The recoil energy of the daughter nucleus from the decay is $E_R^{N'} \lesssim \mathcal{O}(10)\,$eV. The emitted electron is relativistic and does not deposit enough energy in a material to create a persistent damage track. Similarly, $\gamma$-rays lose their energy by scattering off electrons or by electron pair creation, again giving rise to relativistic electrons which do not induce tracks. 

In summary, $\alpha$-decays give rise to a (heavy) recoiling nucleus with energies $10-100$\,keV and an $\alpha$-particle with a few MeV of energy, both of which give rise to potentially observable tracks. $\beta$ and $\gamma$-decays on the other hand give rise to relativistic electrons which leave no persistent damage in materials and recoils of the (heavy) nuclei with energies of order $10\,$eV. Such low-energy recoils give rise to unobservably short damage tracks. Hence, only $\alpha$-decays give rise to potential background events.

Recalling the $^{238}$U decay chain, cf. Eq.~\eqref{eq:Uchain}, and considering the half-lives of the involved nuclei listed in Tab.~\ref{tab:U_halflife} we find that the half-life of the initial $^{238}$U decay is comparable or somewhat larger than the integration time of paleo-detectors. The subsequent decays are much faster; the longest half-life of the nuclei in the decay chain is for $^{234}$U. The accumulated half-life of the decays after $^{222}$Rn not listed in Tab.~\ref{tab:U_halflife} is $< 23\,$yr and the decay chain contains 4 $\alpha$-decays between $^{222}$Rn and the stable $^{206}$Pb. 

Thus, almost all $^{238}$U nuclei which underwent the initial ($^{238}{\rm U} \to {^{234}{\rm Th}} + \alpha$) decay will have decayed further along the decay chain to stable $^{206}{\rm Pb}$. Such events will manifest in the mineral as a sequence of 8 spatially connected $E_R = \mathcal{O}(100)\,$keV recoils of the heavy daughter nuclei in the decay chain accompanied by 8 $\alpha$-tracks. Note that the typical range of an $\alpha$-particle with energies of order MeV is larger than a few $\mu$m in the target minerals of interest.

The characteristic pattern of nuclear recoil tracks can be used to efficiently mitigate $^{238}$U decay events even under the pessimistic assumption that the damage track from the $\alpha$-particles does not create sufficient damage in the target material to be resolved when reading out the material. 

However, due to the relatively long half-life of $^{234}$U, the second $\alpha$-decay in the $^{238}$U decay chain, there will be a population of events in the target sample which has undergone only a single $\alpha$-decay. This background posed a significant problem~\cite{Collar:1995aw,SnowdenIfft:1996zz} in the original attempt to search for DM induced damage tracks in ancient Mica~\cite{SnowdenIfft:1995ke,Engel:1995gw,SnowdenIfft:1997hd,Collar:1994mj}. In~\cite{SnowdenIfft:1995ke} the authors used atomic force microscopy (AFM) to search for damage tracks after cleaving and chemically etching a Mica sample. In Mica, $\alpha$-tracks are not etchable, hence, only the recoil track of the $72\,$keV $^{230}$Th nucleus from the ($^{238}{\rm U} \to {^{234}{\rm Th}} + \alpha$) decay would be detectable. Due to etching the sample and using AFM as read-out which can only scan the surface of the material, the track length resolution in Ref.~\cite{SnowdenIfft:1995ke} was relatively poor. Thus, the characteristic length of the $^{234}$Th induced track could not be used for rejection of that background. However, the advances in read-out technology in the last decades now allow to efficiently mitigate the single-$\alpha$ background.

The number of single-$\alpha$ decays in a target sample is given by
\begin{equation}
 N_{1\alpha}^{238}(t) = N_{238}^0 \frac{\lambda_{238}}{\lambda_{234}-\lambda_{238}} \left( e^{-\lambda_{238} t} - e^{-\lambda_{234} t} \right)\;,
\end{equation}
where $\lambda_{238}$ ($\lambda_{234}$) is the decay constant of $^{238}$U ($^{234}$U) related to the half-life as $\lambda = \ln(2)/T_{1/2}$ and $t$ is the age of the mineral. The initial number of $^{238}$U atoms in the sample, $N_{238}^0$ is given by
\begin{equation}\label{eq:N238_0}
 N_{238}^0 = M_T C^{238} N_A / m_{238}^{\rm mol}\;,
\end{equation}
where $M_T$ is the mass of the target mineral, $C^{238}$ the fraction of $^{238}$U in the target material in weight, $N_A$ the Avogadro constant, and $m_{238}^{\rm mol}$ the molar mass of $^{238}$U. 

For $T_{1/2}^{234} \lesssim t \lesssim T_{1/2}^{238}$, the age of most minerals of interest for paleo-detectors, the number of single-$\alpha$ events per unit target mass is well approximated by
\begin{equation}\begin{split}\label{eq:n_1alpha}
 n_{1\alpha}^{238} & \equiv \frac{N_{1\alpha}^{238}}{M_T} \simeq \frac{N_{238}^0}{M_T} \frac{\lambda_{238}}{\lambda_{234}} \\
 &= 10^{9}\,{\rm kg}^{-1} \left( \frac{C^{238}}{0.01\,{\rm ppb}}\right) \;.
\end{split}\end{equation}

Such large numbers of background events require effective background suppression in order to retain sensitivity to hypothetical DM signals. In principle, there are two options to mitigate the single-$\alpha$ background depending on the read-out scenario discussed in Sec.~\ref{sec:ReadOut}: 1) in the {\it with low-Z tracks} scenario, the coincident detection of an $\alpha$-induced track and the damage track from the recoiling $^{234}$Th nucleus make background suppression trivial. The $\alpha$-particle has a characteristic energy of $4.2\,$MeV, giving rise to tracks longer than $10\,\mu$m in all target material considered. Such tracks are $1-2$ orders of magnitude longer than WIMP-induced tracks. 2) in the {\it without low-Z tracks} scenario, only the track of the $^{234}$Th nucleus is visible. Then, one has to rely on the monochromatic $72\,$keV recoil energy of the $^{234}$Th nuclei, cf. Fig.~\ref{fig:Spectra} for the corresponding track lengths. Note that the finite track length resolution will turn mono-chromatic recoils into a Gaussian spectrum (after reconstruction) with width given by the track length resolution. As we will see in Sec.~\ref{sec:Reach}, the impact on the projected sensitivity of paleo-detectors is negligible since the corresponding narrow feature in track length is easily rejected for high-resolution read-outs such as HIM. For read-out methods with worse resolution such as SAXs, the single-$\alpha$ background depreciates the sensitivity to low mass WIMPs, while the sensitivity to heavier WIMPs is retained.

%*************************
\subsection{Neutrons}\label{sec:Bkg:SF}
A small fraction of the $^{238}$U atoms in the target sample will undergo spontaneous fission (SF) instead of $\alpha$-decays. The fission event itself is an easily rejected background: Spontaneous fission gives rise to two (heavy) daughter nuclei which recoil against each other and subsequently decay to stable nuclei. Such events will give rise to signatures even more spectacular than the 8 $\alpha$-decays in the usual $^{238}$U decay chain discussed above.

However, SF of $^{238}$U nuclei also gives rise to $\sim 2$ fast neutrons with typical energies of $\sim1\,$MeV. Fast neutrons lose their energy predominantly via elastic scattering off target nuclei. The mean free path for fast neutrons in typical target minerals is of the order of a few cm, making it difficult to connect the induced recoil of a target nucleus with the SF event. Furthermore neutrons lose only a small fraction of their energy when elastically scattering off the typically much heavier target nuclei due to the scattering kinematics. Typically, neutrons will undergo $100-1000$ elastic scatterings before losing enough energy to not be able to give rise to nuclear recoils with energies similar to WIMP induced recoils. In conventional direct detection experiments, such multiple scatterings are used to reject neutron induced recoil events. However, since paleo-detectors have no timing information, such an approach cannot be used for background rejection. 

The number of spontaneous fission events per unit target mass in a target mineral of age $t$ from $^{238}$U contamination is given by
\begin{equation}
 n_{\rm SF}^{238} = \frac{N_{238}^0}{M_T} \left( 1 - e^{-\lambda_{238} t} \right) \frac{T_{1/2}^{238}}{T_{1/2}^{238;\,{\rm SF}}} \;,
\end{equation}
with the initial number of $^{238}$U atoms in the sample, $N_{238}^0$ given by Eq.~\eqref{eq:N238_0}, $M_T$ the mass of the target mineral, $\lambda_{238} = \ln(2)/T_{1/2}^{238}$ the decay constant of $^{238}$U, and $T_{1/2}^{238}$ ($T_{1/2}^{238;\,{\rm SF}}$) the (SF) half-life of $^{238}$U reported in Tab.~\ref{tab:U_halflife}. For mineral ages short compared to the $^{238}$U half-life, the number of SF events is well approximated by
\begin{equation}
 n_{\rm SF}^{238} \simeq 2 \times 10^6\,{\rm kg}^{-1} \left( \frac{C^{238}}{0.01\,{\rm ppb}}\right) \left( \frac{t}{1\,{\rm Gyr}} \right) \;.
\end{equation}

Fast neutrons are also produced in so-called $(\alpha,n)$-reactions, emission of fast neutrons from interactions of $\alpha$-particles with heavier nuclei. Although only a small fraction of $\alpha$-particles will lead to neutron emission, $(\alpha,n)$-reactions yield a sizeable contribution to the neutron flux in the target since the number of $\alpha$-particles is $\sim 10^7$ times larger than the number of neutrons from SF events. Depending on the composition of the target material, the neutron spectrum can either be dominated by neutrons from SF events or by neutrons from $(\alpha,n)$ interactions. Generally, the lighter the target nuclei are, the more relevant the $(\alpha,n)$ contribution becomes. However, the strong dependence of the $(\alpha,n)$ cross section on the nuclear structure of the target isotopes makes general statements difficult.

We use the \texttt{SOURCES-4A} code~\cite{sources4a:1999} to calculate the neutron spectra from SF of all nuclei in the $^{238}$U decay chain and $(\alpha,n)$ reactions induced by $\alpha$-particles from the $^{238}$U decay chain. From these spectra, we calculate the induced nuclear recoil spectra using the neutron-nucleon cross sections tabulated in the \texttt{JANIS4.0} database~\cite{Soppera:2014zsj}.\footnote{We use values from \texttt{TENDL-2017}~\cite{Koning:2012zqy,Rochman:2016,Sublet:2015,Fleming:2015} for neutron-nucleon cross sections.} We take only elastic neutron-nucleon scattering into account. Thus, the obtained background is a conservative estimate: including additional processes such as inelastic scattering, neutron absorption, or $(n,\alpha)$ processes lowers the background, since neutrons lose a larger fraction of their energies in such interactions than in elastic scatterings, or are absorbed. 

Due to the scattering kinematics, neutrons lose only a small fraction of their energy when elastically scattering off heavy nuclei. However, when scattering off light targets, the energy transfer is much more efficient: On average, a fast neutron from SF or $(\alpha,n)$-reactions will give rise to $\sim 4$ ($\sim 7$) nuclear recoils with $E_R \gtrsim 10\,$keV ($E_R \gtrsim 1\,$keV) when scattering off hydrogen with $m_T \sim 1\,$GeV. When scattering off nuclei with $m_T \sim 10\,$GeV, the same neutrons give rise to $\sim 20$ ($\sim 50$) recoils with energies $E_R \gtrsim 10\,$keV ($E_R \gtrsim 1\,$keV). For even heavier nuclei, with $m_T \sim 100\,$GeV, fast neutrons induce $\sim 30$ ($\sim 200$) recoils with energies $E_R \gtrsim 10\,$keV ($E_R \gtrsim 1\,$keV). Because of the kinematic match of the neutron and proton (i.e. H nuclei) masses and because the neutron-hydrogen elastic scattering cross section is large compared to those of most heavier nuclei, fast neutrons will scatter efficiently off hydrogen in a target and lose a large fraction of their energy in each interaction. This leads to a large reduction of the number of energetic neutron induced nuclear recoils, in particular recoils of nuclei heavier than hydrogen, even if hydrogen comprises only a relatively small fraction of the target molecules, as can be seen by comparing the left and right panels of Fig.~\ref{fig:Spectra}. In addition, depending on the target material and read-out method, the hydrogen recoils themselves may not give rise to observable tracks, cf. the discussion in Sec.~\ref{sec:ReadOut}.

On the other hand, $(\alpha,n)$ cross sections are typically large for the lightest target nuclei with $Z \geq 3$. Thus, target minerals containing lithium or beryllium are not well suited for paleo-detectors.

We show the track length spectra induced by neutrons in Fig.~\ref{fig:Spectra} together with the WIMP-induced spectra. Trivially, the neutron induced background is lower in materials with lower concentration of $^{238}$U. However, the difference between target minerals with and without hydrogen is much larger, suppressing the neutron induced background by more than two orders of magnitudes between otherwise similar target minerals.

\begin{figure*}
 \includegraphics[width=0.49\linewidth]{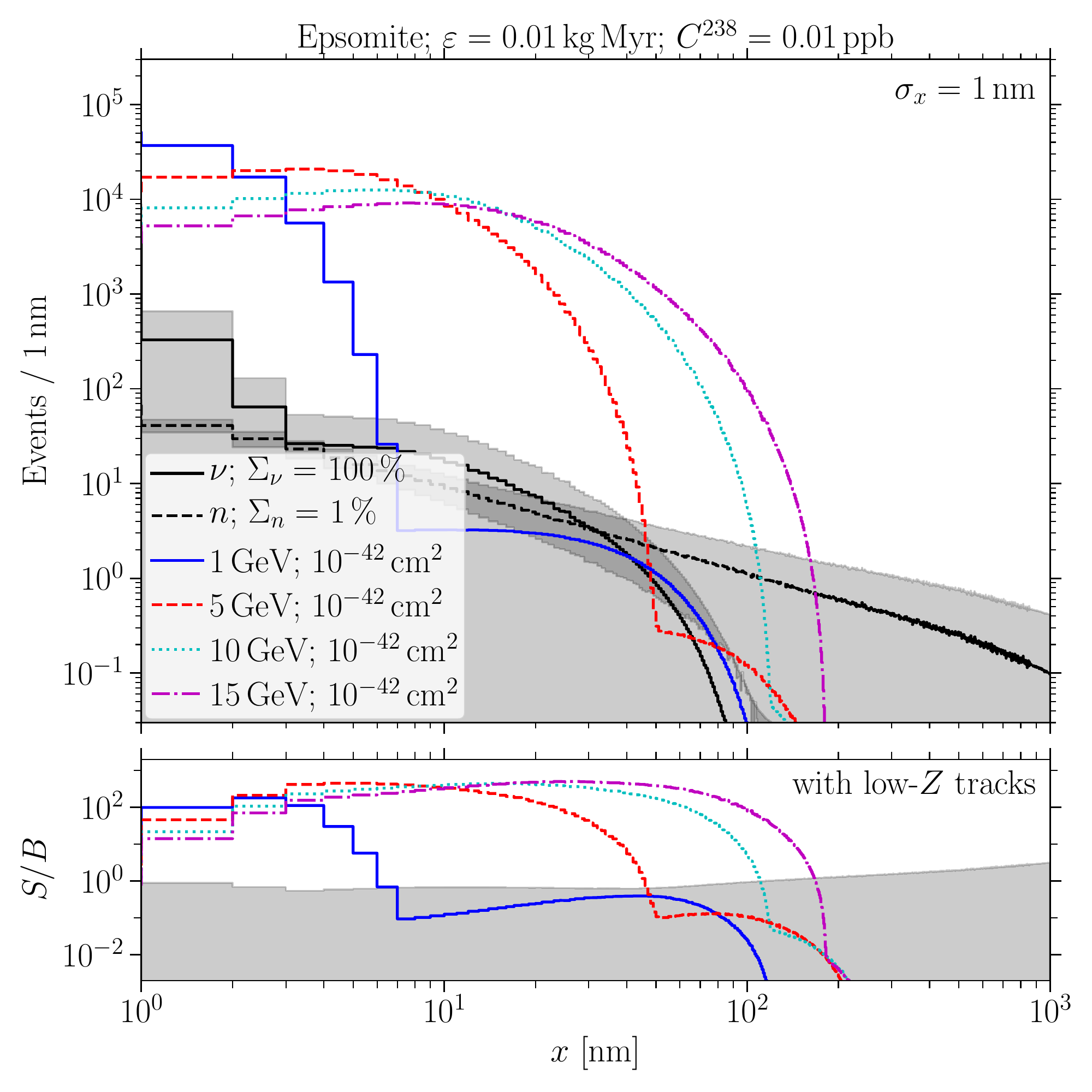}
 \includegraphics[width=0.49\linewidth]{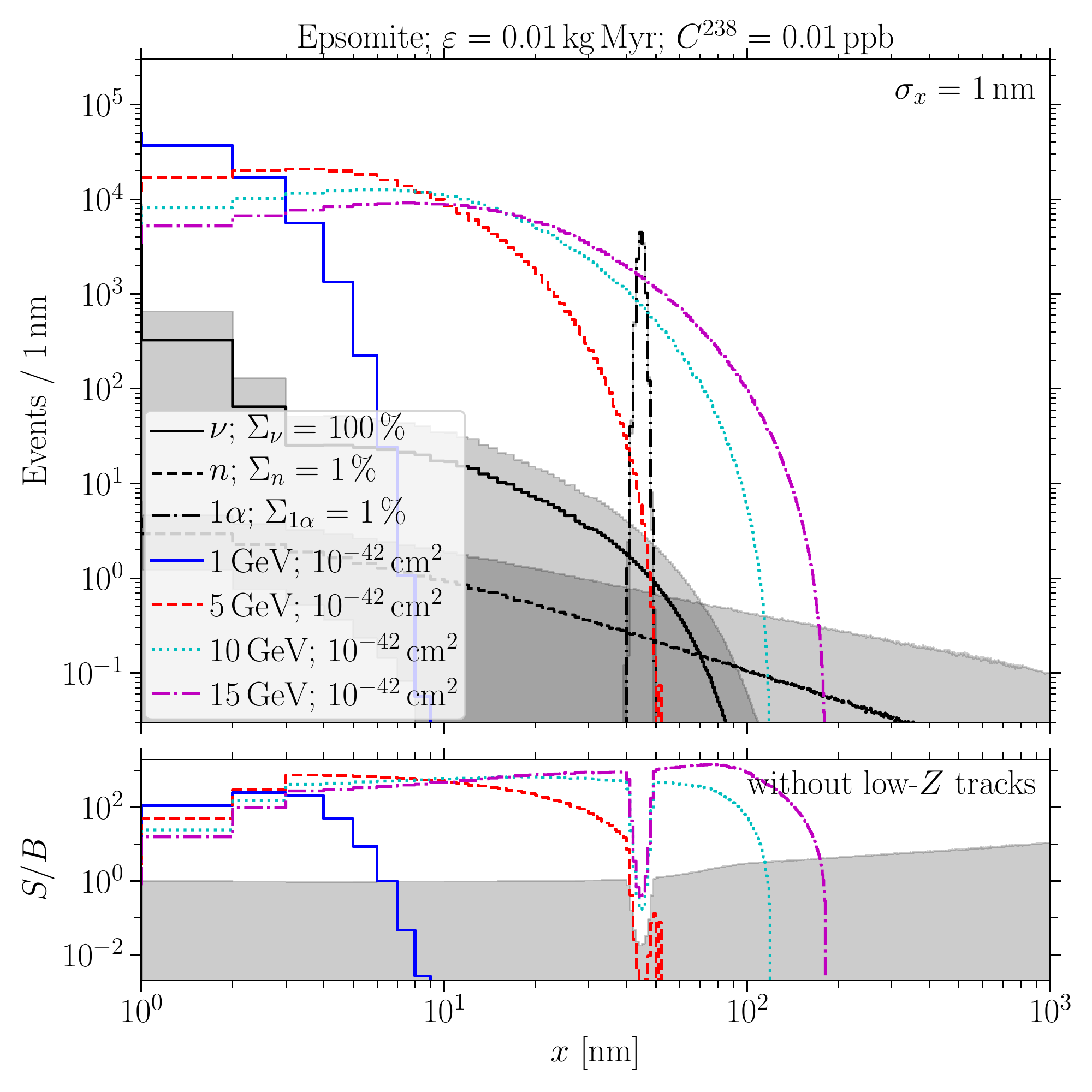}
 \caption{Binned recoil spectra (upper panels) and ratio of signal ($S$) to background ($B$) events per bin (lower panels) in Epsomite [\ccEps], assuming an exposure of $0.01\,$kg\,Myr and a $^{238}$U concentration of $C^{238} = 0.01\,$ppb in weight. The colored lines are for WIMPs with (5, 10, 15, 50)\,GeV mass assuming a WIMP-nucleon cross section of $\sigma_n^{\rm SI} = 10^{-42}\,$cm$^2$ as indicated in the legend. {\it Upper panels}: The solid, dashed, and dash-dotted black lines are for the neutrino ($\nu$), neutron ($n$), and single-$\alpha$ ($1\alpha$) induced background spectra, respectively. The gray shaded bands around the respective lines indicate the statistical and systematic errors, added in quadrature, of the respective backgrounds. Note that the uncertainty bands of multiple backgrounds overlap where the shaded area appears darker. {\it Lower panels:} Ratio of signal ($S$) to total background ($B$) events per bin. The gray shaded area indicates the relative uncertainty of the total background per bin, i.e. the ratio of the (quadratic sum of the statistical and systematic errors of all components)/(sum of all background events) per bin. The signal-to-noise ratio, as defined in Eq.~\eqref{eq:SNR}, per bin is obtained by dividing $S/B$ for the respective signal with the relative uncertainty of the background. The left panels show spectra calculated {\it with low-$Z$ tracks}, i.e. under the assumption that tracks from low-$Z$ nuclei, including hydrogen, are visible. Thus, the single-$\alpha$ induced background can be rejected based on the appearance of the $\alpha$-tracks. The spectra in the right panels are calculated {\it without low-$Z$ tracks}. Thus, we assume that low-$Z$ nuclei such as hydrogen and helium ($\alpha$-particles) do not leave visible tracks. Hence, the monochromatic $^{234}$Th recoils from the single-$\alpha$ background appear as broadened spectra due to the finite track length resolution.} 
 \label{fig:SNR_Eps_LM}
\end{figure*}

%*************************
\subsection{Neutrinos}
Neutrinos emitted in the Sun, supernovae explosions, and cosmic ray interactions in the atmosphere can coherently scatter off the target nuclei giving rise to nuclear recoils; the same process which gives rise to the {\it neutrino floor} for DD experiments~\cite{Billard:2013qya}. The differential recoil spectrum per unit target mass induced by neutrinos is given by~\cite{Billard:2013qya,OHare:2016pjy}
\begin{equation}
 \left(\frac{dR}{dE_R}\right)_T = \frac{1}{m_T} \int_{E_\nu^{\rm min}} dE_\nu \, \frac{d\sigma}{dE_R} \frac{d\Phi_\nu}{dE_\nu}\;,
\end{equation}
where $E_\nu^{\rm min} = \sqrt{m_T E_R/2}$ is the minimum neutrino energy required to induce a nuclear recoil with energy $E_R$, similar to the maximum momentum transfer in Eq.~\eqref{eq:diffxsec}. The differential cross section for coherent neutrino-nucleus scattering is
\begin{equation}
 \frac{d\sigma}{dE_R}(E_R,E_\nu) = \frac{G_F^2}{4\pi} Q_W^2 m_T \left( 1 - \frac{m_T E_R}{2E_\nu^2} \right) F^2(E_R)\;,
\end{equation}
where $G_F$ is the Fermi coupling constant, [$Q_W \equiv \left(A_T - Z_T\right) - \left(1-4\sin^2\theta_W\right) Z_T$] with the weak mixing angle $\theta_W$, and $F(E_R)$ is the nuclear form factor. We take the neutrino flux $d\Phi_\nu/dE_\nu$ from Ref.~\cite{OHare:2016pjy}; for $E_\nu \lesssim 20\,$MeV the neutrino flux is dominated by solar neutrinos, for $20\,{\rm MeV} \lesssim E_\nu \lesssim 30\,$MeV by the diffuse supernova neutrino background (DSNB), and for larger energies $E_\nu \gtrsim 30\,$MeV by atmospheric neutrinos. The nuclear recoil spectrum due to the neutrino background is converted to an ionization track length spectrum analogously to the WIMP induced recoil spectra; we show the resulting spectrum together with those induced by WIMPs and neutrons in Fig.~\ref{fig:Spectra} for a selection of target materials.

%*************************
\subsection{Background Discussion}\label{sec:Bkg_discussion}

\begin{figure*}
 \includegraphics[width=0.49\linewidth]{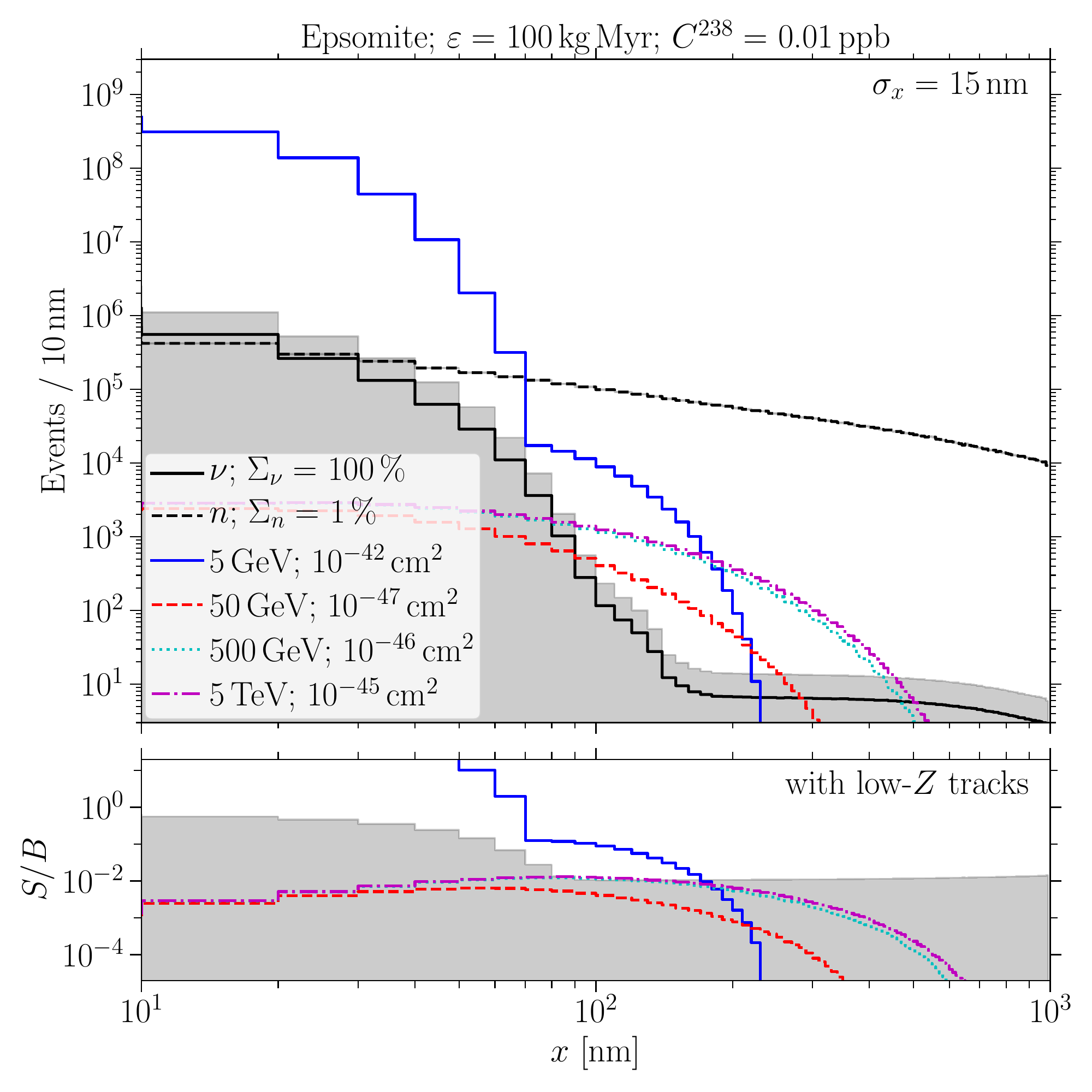}
 \includegraphics[width=0.49\linewidth]{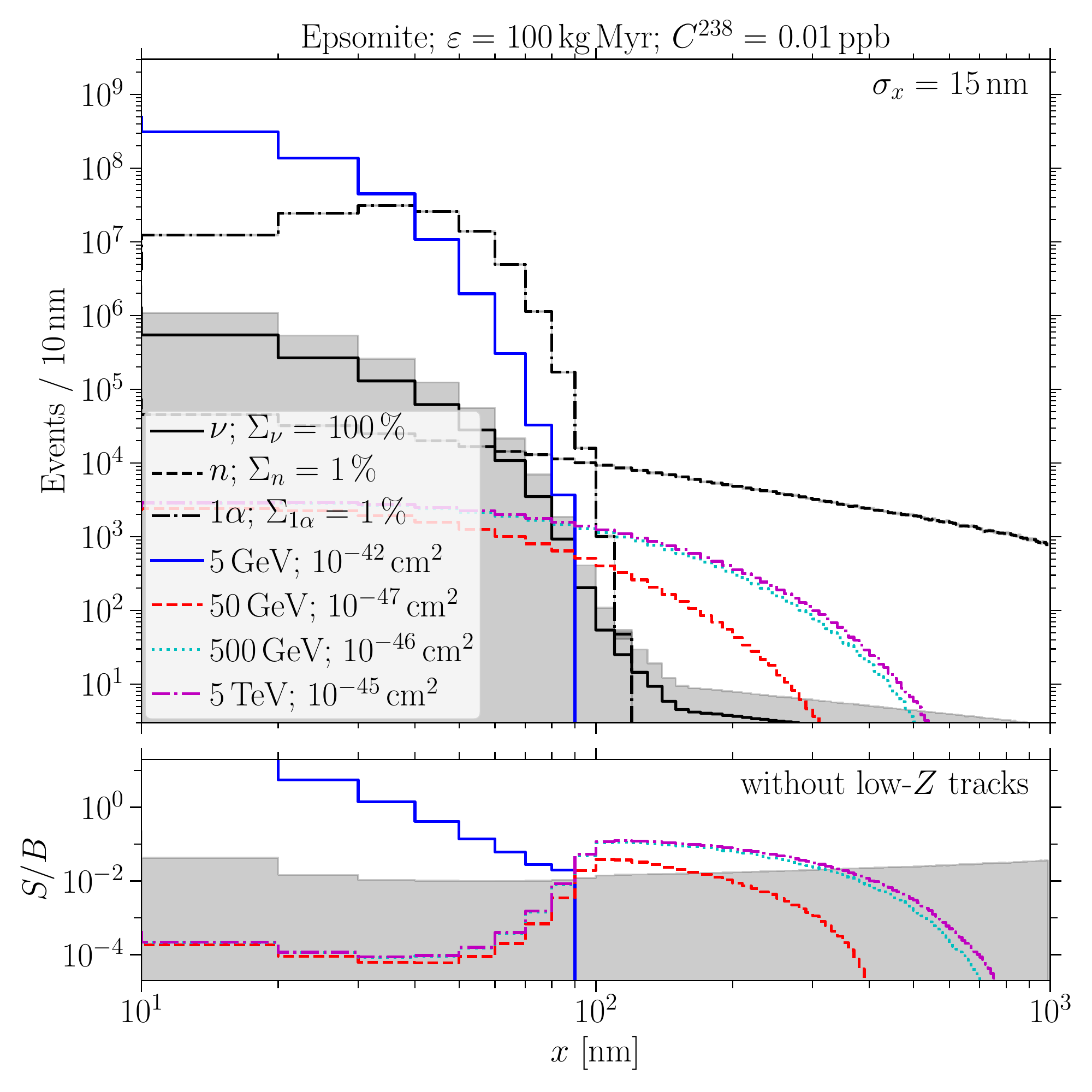}
 \caption{Same as Fig.~\ref{fig:SNR_Eps_LM} but for an exposure of $100\,$kg\,Myr and assuming a track length resolution of $\sigma_x = 15\,$nm. WIMP signal are shown for masses of (5, 50, 500, 5000)\,GeV and assuming WIMP-nucleon cross sections of $(10^{-42},\;10^{-47},\;10^{-46},\;10^{-47})\,$cm$^2$, respectively. Note that the bin-width is $10\,$nm instead of $1\,$nm in Fig.~\ref{fig:SNR_Eps_LM} and that the scales of the axes are different.}
 \label{fig:SNR_Eps_HM}
\end{figure*}

In Fig.~\ref{fig:Spectra} we compare the track length spectra induced by the respective backgrounds with those induced by WIMPs. A more detailed comparison including the uncertainty of the background and the effects of finite track length resolution on the spectra is shown for Epsomite in Fig.~\ref{fig:SNR_Eps_LM} for low-mass WIMPs and in Fig.~\ref{fig:SNR_Eps_HM} for heavier WIMPs. In the upper panels of these figures, we indicate the neutrino, neutron, and single $\alpha$-decay induced backgrounds with the solid, dashed, and dashed-dotted black lines, respectively. The shaded bands around the lines show the systematic and statistical errors, added in quadrature, of the respective backgrounds. Colored lines show the WIMP-induced spectra for a variety of WIMP masses and benchmark WIMP-nucleon scattering cross sections, cf. the captions of Figs.~\ref{fig:SNR_Eps_LM} and~\ref{fig:SNR_Eps_HM}. In addition, the bottom panels show the ratio of the number of signal to background events for each WIMP benchmark together with the relative uncertainty of the total background.

The relevant background quantity is not the total number of events, but the uncertainty of the number of background events in the signal region. Here, we assume that backgrounds induced by neutrinos suffer from much larger uncertainties than radioactivity induced backgrounds. The main uncertainty on the neutrino induced backgrounds stems from the normalization of the neutrino fluxes. Today's fluxes are known to an accuracy of a few percent for solar neutrinos, while the DSNB and atmospheric neutrino fluxes are only known within some tens of percent~\cite{OHare:2016pjy}. However, paleo-detectors would measure the time-integrated neutrino induced background over a period of as large as a billion years, entailing further uncertainties from extrapolating the current neutrino fluxes over such time scales. As in Ref.~\cite{Baum:2018tfw} we account for such uncertainties by assuming a relative systematic error of $\Sigma_\nu = 100\,\%$ for the number of neutrino induced background events. Note that a spectral analysis would allow one to significantly reduce this error since such an analysis would use a control region dominated by neutrino induced events to measure the normalization of the neutrino induced background~\cite{PAPER3}. It should be kept in mind however that the normalization of the neutrino induced backgrounds from the different components of the solar neutrino fluxes, DSNB, and atmospheric neutrinos may fluctuate independently such that each component must be measured individually. 

The normalization of the radioactive backgrounds on the other hand is determined solely by the initial concentration of radioactive materials in the target mineral and the exposure time. There is no external source which may fluctuate with time as for the neutrino induced backgrounds. Hence, the normalization can be predicted much better, e.g. from measuring the number of full $^{238}$U decay chains in the target mineral. We assume a systematic uncertainty of 1\,\% on the single-$\alpha$ and neutron induced backgrounds.

For the assumed benchmark $^{238}$U concentration of $C^{238} = 0.1\,$ppb ($C^{238} = 0.01\,$ppb) for UBR (ME) minerals, we can identify two (broadly speaking) background regimes. For tracks shorter than some tens of nm, the background uncertainty is dominated by recoils induced by solar neutrinos. For longer tracks, the neutrino induced background is dominated by DSNB and atmospheric neutrinos, both of which have much smaller fluxes than solar neutrinos. Thus, for tracks longer than some tens of nm, neutron induced events are the dominant contribution to the background uncertainty. Note that the value of the track length below (above) which solar neutrinos (neutrons) dominate the background budget is target dependent.

The single-$\alpha$ background plays a role only if $\alpha$-tracks cannot (reliably) be reconstructed by the chosen read-out method. However, the impact on the signal-to-noise ratio remains small as can be seen by comparing the left and right bottom panels of Figs.~\ref{fig:SNR_Eps_LM} and~\ref{fig:SNR_Eps_HM}, respectively.
 
%*************************
\section{Mineral Optimization}\label{sec:MinOpt}
%*************************
As discussed in Sec.~\ref{sec:Bkg}, one of the major sources of backgrounds in paleo-detectors is decays of radioactive contaminants, in particular $^{238}$U. In order to suppress this background, materials with as low $^{238}$U concentration as possible must be chosen as target materials. The typical uranium concentration in the Earth's crust is of the order of parts per million (ppm) in weight, which would lead to unacceptably large backgrounds. Much lower concentrations of uranium are found in the Earth's mantle and in seawater. Thus, promising target materials for paleo-detectors are {\it ultra-basic rocks} (UBRs), formed in the Earth's mantle, and {\it marine evaporites} (MEs), formed at the bottom of evaporating oceans.\footnote{Recall that for brevity, we refer to minerals in marine evaporite deposits as ``marine evaporites'' and to minerals in ultra-basic rocks as ``ultra-basic rocks''.} The most common example of an UBR is Olivine [\ccOli], and the most common MEs are Halite (NaCl) and Gypsum [\ccGyp]. Note that further purification of the materials may arise from chemical expulsion of contaminants during the growth of crystals. However, the effect of such purification can not be quantified in general, see e.g.~\cite{Adams:1959}.\footnote{Once obtained, concentrations of radioactive trace elements in target samples of interest for paleo-detectors can be measured reliably to levels as low as $\sim 10^{-15}$ in weight using e.g. inductively coupled plasma mass spectroscopy.} Here, we use benchmark values for the $^{238}$U concentrations of $0.01\,$ppb (parts per billion) in weight for MEs and $0.1\,$ppb for UBRs. 

As discussed in Sec.~\ref{sec:Bkg:SF}, neutron induced backgrounds are further suppressed by the presence of hydrogen in the material since hydrogen is an effective moderator of fast neutrons. Suppression of the neutron induced background is particularly relevant to maintain sensitivity to WIMPs with mass $m_\chi \gtrsim 10\,$GeV. Examples of UBRs containing hydrogen are Nchwaningite [\ccNch] and Nickelbischofite [\ccNic], while MEs containing hydrogen are e.g. Gypsum [\ccGyp] or Epsomite [\ccEps].

Further gains in sensitivity can be made by choosing materials which are optimal for particular ranges of WIMP masses or WIMP-nucleus interaction types. For example, for low-mass WIMPs one would preferentially use minerals with low mass density and low-mass target nuclei. In such targets, tracks are relatively long, such that recoils induced by low-mass WIMPs are more easily read out. On the other hand, light nuclei have comparatively large $(\alpha,n)$ cross sections, rendering neutron-induced backgrounds in such targets challenging.

For probing SI interactions of WIMPs with masses $m_\chi \gtrsim 10\,$GeV, it is advantageous to have as many heavy target nuclei in the target material as possible since the WIMP-nucleus cross section is coherently enhanced with the number of nucleons, $\sigma_T^{\rm SI} \propto A_T^2$. However, MEs and UBRs containing {\it both} hydrogen and nuclei heavier than Ni are quite rare.

For probing SD interactions, the target material must contain nuclei with large nuclear spin. Interesting target elements for WIMP-proton SD interactions are for example H, B, F, Na, Al, K, or Mn. Note that out of the first two, hydrogen is only useful as a target in the {\it with low-$Z$ tracks} scenario. Boron is a good target for WIMP-proton SD scattering, however, it also leads to relatively large backgrounds from ($\alpha,n$)-interactions. Examples of UBRs containing such nuclei are Nchwaningite [\ccNch] and Phlogopite [\ccPhl], while Borax [\ccBor] and Mirabilite [\ccMir] are examples of MEs.

The situation is worse for WIMP-neutron SD scattering, as natural targets do not usually contain sizeable fractions of target isotopes with unpaired neutrons. The most interesting target elements are Mg, Si and Zr. Note that O offers some sensitivity to WIMP-neutron scattering as well, although the abundance of the relevant isotope $^{17}$O is small. Examples of UBR targets are Baddeleyite [\ccBad] and Phlogopite [\ccPhl], Example of MEs are Cattiite [\ccCat] and Epsomite [\ccEps].

Besides the materials mentioned above and on which we will focus in this work, other classes of materials may be promising for particular applications. For example, natural plastics such as Evenkite (C$_{24}$H$_{50}$) are excellent SSTDs. Since they are comprised of relatively light target nuclei, natural plastics would be interesting targets for low-mass WIMPs $m_\chi \lesssim 10\,$GeV in particular. However, little is known about typical uranium concentrations in natural plastics such that we do not consider them as targets in this work. Another interesting material is diamond, samples of which can be extremely radiopure~\cite{KRAMERS197958}. However, the diamond sample would have to be much larger than the few cm mean free path of fast neutrons in order for the sample to not be contaminated by nuclear recoils induced by neutrons originating from the surrounding material. In reverse, the background of neutron induced recoils may be suppressed in general if a target mineral of linear dimensions smaller than a few cm is embedded in a uranium-poor environment, e.g. ice or pure salt deposits. In such a situation, the neutrons originating within the sample would mostly scatter outside of the target volume, while the average neutron flux through the sample would be reduced due to the lower radioactivity in the surrounding material. 

%*************************
\section{Sensitivity Estimation}\label{sec:Reach}
%*************************
\begin{figure*}
 \includegraphics[width=0.49\linewidth]{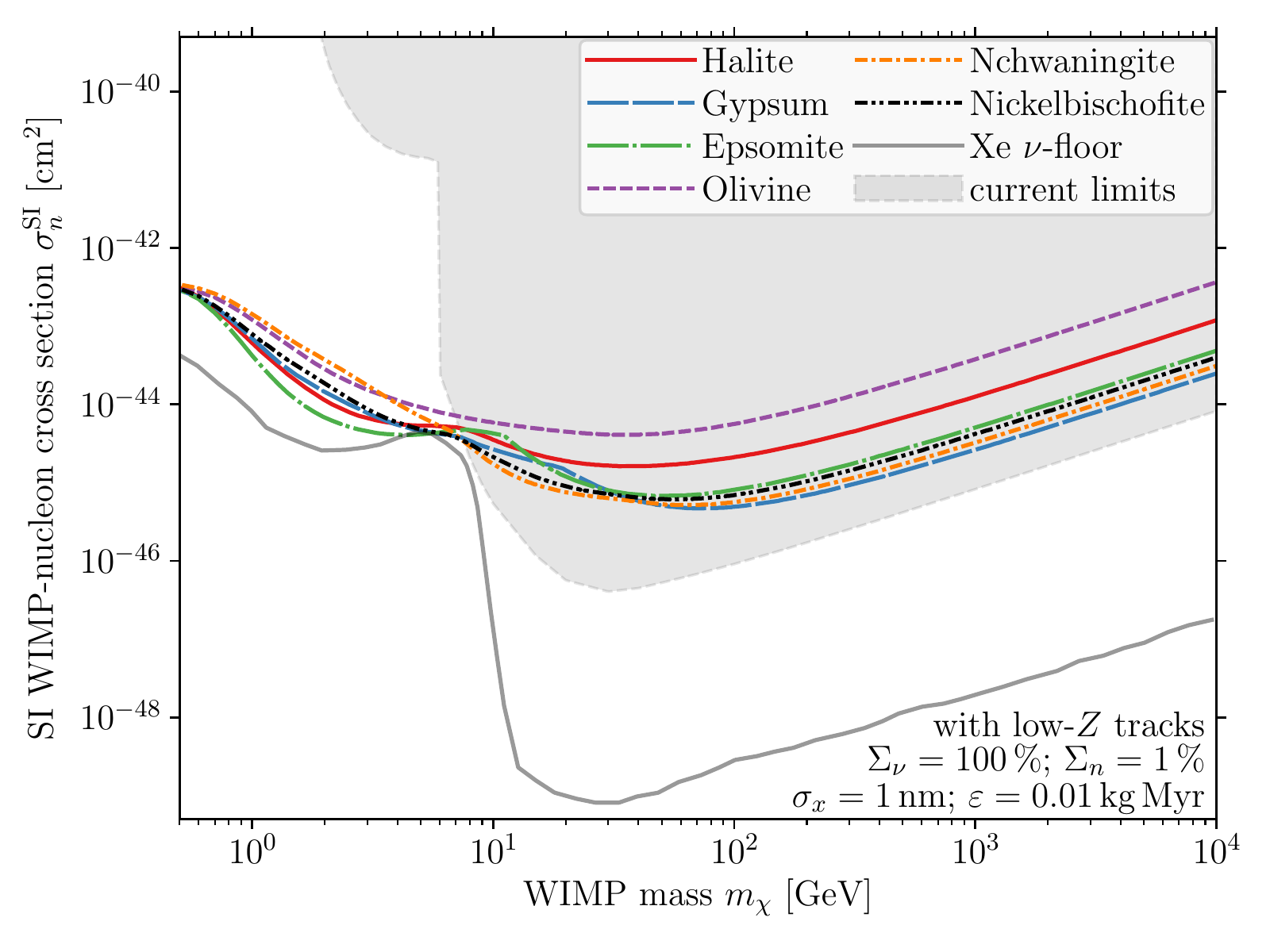}
 \includegraphics[width=0.49\linewidth]{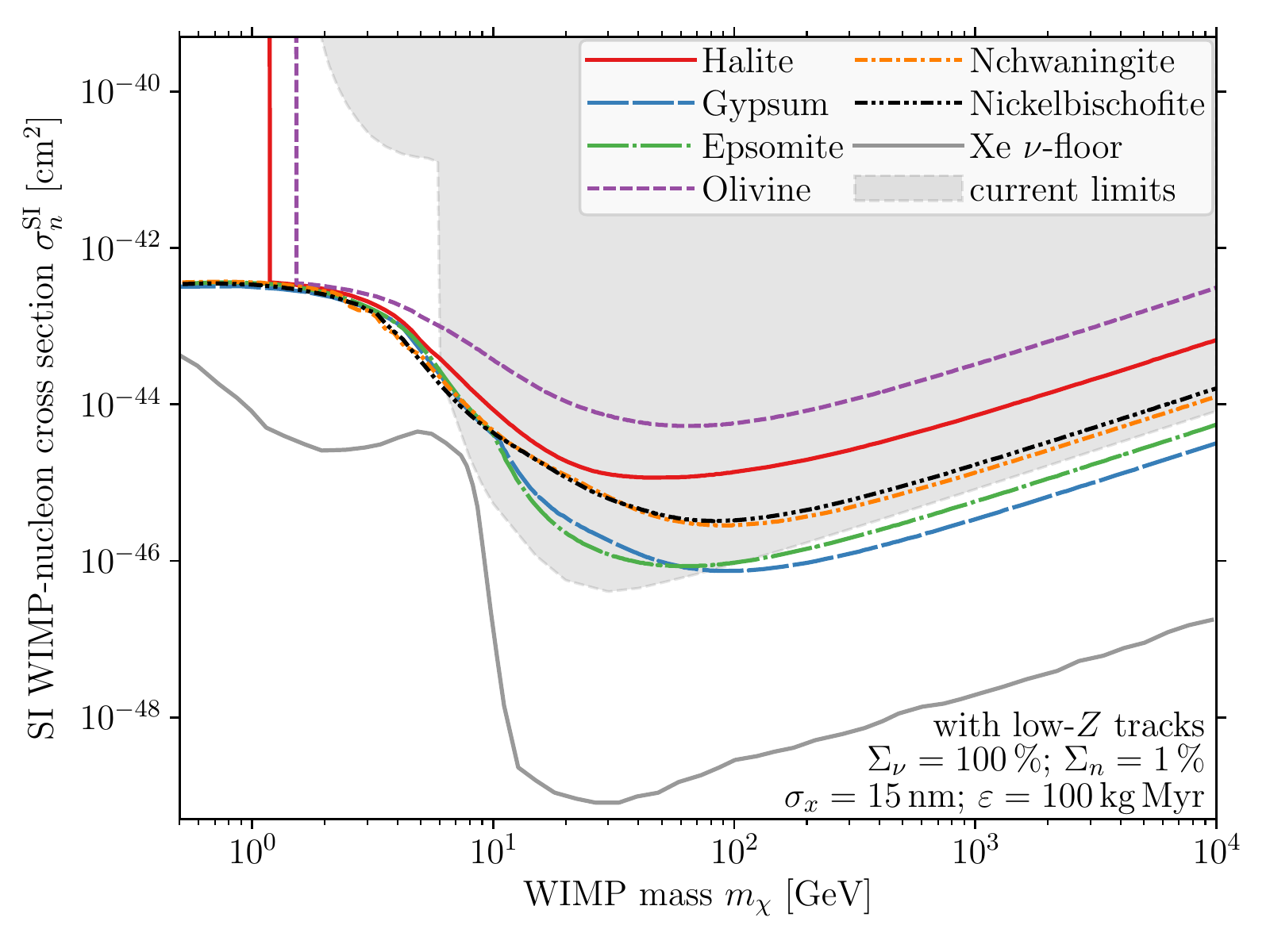}

 \includegraphics[width=0.49\linewidth]{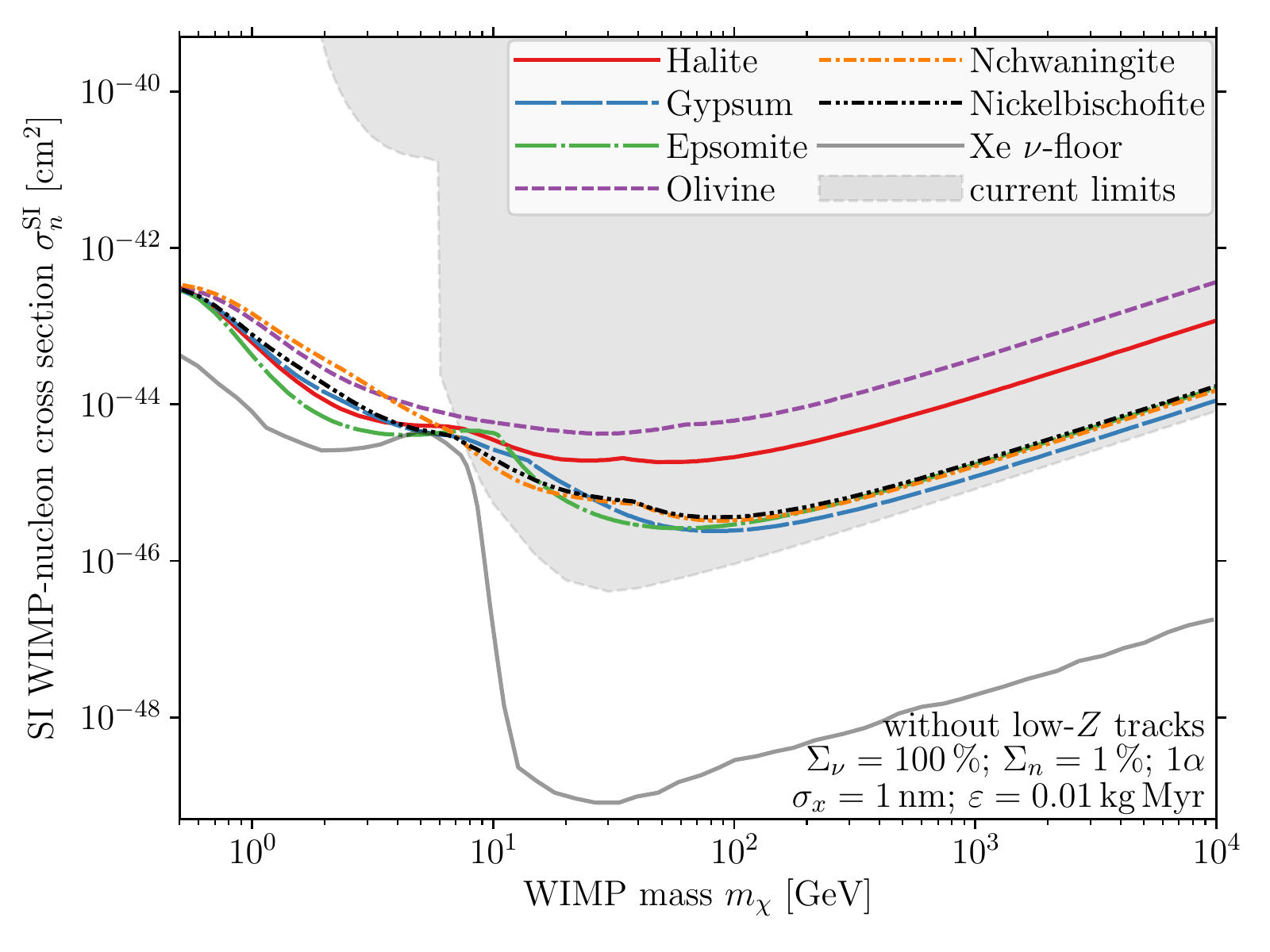}
 \includegraphics[width=0.49\linewidth]{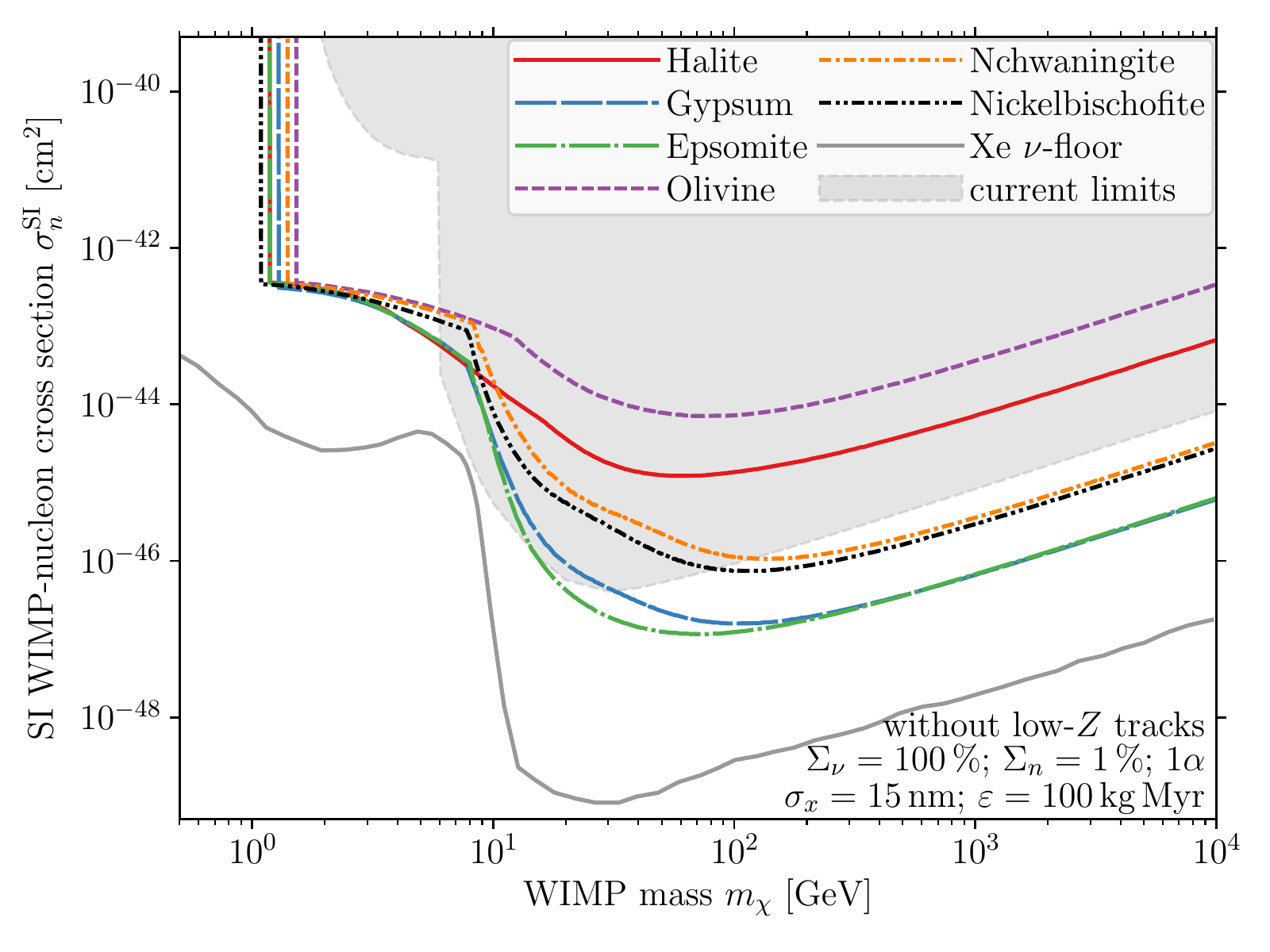}
 \caption{Sensitivity projection for spin-independent (SI) WIMP-nucleon scattering cross sections. The different colors are for different target materials: Three examples of MEs, Halite (\ccHal), Gypsum [\ccGyp], and Epsomite [\ccEps] and three examples of UBRs, Olivine [\ccOli], Nchwaningite [\ccNch], and Nickelbischofite [\ccNic]. For reference, the light gray line indicates the conventional neutrino floor for Xe direct detection experiments~\cite{Cushman:2013zza}. The shaded area shows current direct detection limits~\cite{Angloher:2015ewa,Agnese:2017jvy,Petricca:2017zdp,Aprile:2018dbl}. For the left (right) panels we assume a track length resolution of $\sigma_x = 1\,$nm ($\sigma_x = 15\,$nm) and an exposure of $\varepsilon = 0.01\,$kg\,Myr ($\varepsilon = 100\,$kg\,Myr). The top panels are for the {\it low-$Z$ tracks} scenario where we include tracks from all nuclei. The bottom panels are for the {\it without low-$Z$ tracks} scenario, for which we assume only tracks with $Z \geq 3$ to give rise to reconstructible tracks. We use $^{238}$U concentrations of $C^{238}=0.01\,$ppb in weight for MEs and $C^{238}=0.1\,$ppb for UBRs.}
 \label{fig:SI_projection}
\end{figure*}

From the track length spectra for the signal and the different background components discussed in Secs.~\ref{sec:Sig} and~\ref{sec:Bkg}, respectively, we estimate the sensitivity of paleo-detectors using a simple cut-and-count analysis.

For each WIMP mass hypothesis, we account for the finite track length resolution by sampling the spectra and smearing the track length of each event as
\begin{equation}
 x^{\rm smear} = x^{\rm true} + \Delta_x(\sigma_x) \;,
\end{equation}
where $x^{\rm smear}$ ($x^{\rm true}$) is the smeared (true) track length and $\Delta_x$ a random number with dimensions of length drawn from a normal distribution with standard deviation $\sigma_x$ given by the track length resolution of the respective read-out method. 

We then assume a signal region
\begin{equation}
 x_{\rm min} < x^{\rm smear} < x_{\rm max} \;,
\end{equation}
and count the number of signal (background) events $S$ ($B$) in that signal window. The signal-to-noise ratio is calculated as
\begin{equation} \label{eq:SNR}
 {\rm SNR} = \frac{S}{\sqrt{B_\nu + \Sigma_\nu^2 B_\nu^2 + B_{\rm SF} + \Sigma_{\rm SF}^2 B_{\rm SF}^2 + \ldots}} \,,
\end{equation}
where $\Sigma_i$ is the relative systematic error of the background $i$ and the ``$\ldots$'' indicate additional background components, e.g. the single-$\alpha$ events.

The number of signal events is proportional to the WIMP-nucleon interaction cross section. For each signal region satisfying
\begin{equation}
 \sigma_x/2 \leq x_{\rm min} \leq x_{\rm max} - 2 \sigma_x\;, \quad x_{\rm max} \leq 10^3\,{\rm nm} \,,
\end{equation}
we find the smallest interaction cross section for which
\begin{equation}
 {\rm SNR} > 3 \quad {\rm and} \quad S \geq 5 \,,
\end{equation}
thus finding the signal region for each mass hypothesis yielding the best projected sensitivity in terms of the smallest WIMP-nucleon cross sections which could be probed.

In the remainder of this section, we present sensitivity projections for different benchmark assumptions: We use a high resolution ($\sigma_x = 1\,$nm), low throughput ($\varepsilon = 0.01\,$kg\,Myr) scenario, and a low resolution ($\sigma_x = 15\,$nm), large throughput ($\varepsilon = 100\,$kg\,Myr) scenario. The first scenario may for example be realized by reading out 10\,mg of a $1\,$Gyr old sample with Helium Ion beam Microscopy (HIM). The latter scenario could be realized by reading out 100\,g of a 1\,Gyr old sample with Small Angle X-ray scattering (SAXs); see Sec.~\ref{sec:ReadOut} for a discussion of read-out methods. For both scenarios we discuss sensitivity projections both under the assumption that all nuclei give rise to reconstructible tracks ({\it with low-$Z$ tracks}) and under the assumption that only nuclei with $Z \geq 3$ yield reconstructible tracks ({\it without low-$Z$ tracks}). In particular, in the latter case we assume that hydrogen nuclei and $\alpha$-particles (He nuclei) do not give rise to reconstructible tracks. We note that our analysis only extends down to $m_\chi \simeq 0.5 \,$GeV since WIMPs with mass any lower would not give rise to a significant number of nuclear recoil tracks $\gtrsim 1 \,$nm in any of the target minerals considered here.

As discussed in Sec.~\ref{sec:Bkg_discussion}, we assume a relative systematic error of $\Sigma_\nu = 100\,\%$ for the neutrino induced background. For the backgrounds induced by radioactivity, in particular the neutron and single-$\alpha$ decay induced backgrounds, we assume a relative systematic error of $\Sigma_n = \Sigma_{1\alpha} = 1\,\%$. 

%*************************
\subsection{Spin-Independent Interactions} \label{sec:SIresults}

In Fig.~\ref{fig:SI_projection} we show sensitivity projections for Spin-Independent (SI) WIMP-nucleus interactions in 6 different target minerals. We pick three examples of MEs: The two very common evaporites Halite (\ccHal) and Gypsum [\ccGyp] and the less common evaporite Epsomite [\ccEps]. Halite is an example for an evaporite not containing hydrogen while Gypsum and Epsomite do. Likewise, we choose three examples of UBRs: The very common Olivine [\ccOli], and two minerals containing hydrogen, Nchwaningite [\ccNch] and Nickelbischofite.

Comparing the different panels of Fig.~\ref{fig:SI_projection}, we can first note, that in general, for low-mass WIMPs with masses $m_\chi \lesssim 10\,$GeV good spatial resolution is crucial, while for heavier WIMPs large exposure is more relevant. This is because lighter WIMPs give rise to less energetic recoils and hence shorter tracks than heavier WIMPs. As discussed in Sec.~\ref{sec:Bkg_discussion}, for the assumed $^{238}$U concentrations of $C^{238} = 0.01\,$ppb ($C^{238} = 0.1\,$ppb) for MEs (UBRs), the dominant background source for WIMPs with $m_\chi \lesssim 10\,$GeV are solar neutrinos. Hence, for this mass regime, the sensitivity of MEs and UBRs as well as targets containing or not containing hydrogen is comparable. We find that the typical sensitivity extends to SI WIMP-nucleon cross sections as small as $\sigma_n^{\rm SI} \sim 10^{-43}\,$cm$^2$ for WIMP masses $m_\chi \sim 1\,$GeV and to cross sections approximately one order of magnitude smaller for WIMPs with masses $m_\chi \sim 10\,$GeV.

For heavier WIMPs with masses $m_\chi \gtrsim 10\,$GeV we find appreciable differences between targets containing or not containing hydrogen as well as between MEs and UBRs. This is because for such WIMP masses, the background is dominated by neutron induced recoils. Since the neutron flux is proportional to the $^{238}$U concentration in the target material, MEs are more promising targets assuming that uranium concentrations in MEs are one order of magnitude lower than in UBRs. Further, the sensitivity is significantly increased in minerals containing hydrogen. If tracks from low-$Z$ nuclei, in particular hydrogen, are reconstructed during read-out and cannot be differentiated from tracks from heavier nuclei, only MEs containing hydrogen have projected sensitivities beyond current limits for $m_\chi \gtrsim 10\,$GeV, cf. the top right panel of Fig.~\ref{fig:SI_projection}. If hydrogen nuclei leave no reconstructible tracks or if hydrogen tracks can reliably be differentiated from tracks from heavier nuclei, both UBRs and MEs have projected sensitivities better than current limits, cf. the bottom right panel of Fig.~\ref{fig:SI_projection}. The sensitivity of MEs containing hydrogen such as Gypsum or Epsomite are approximately a factor $10-100$ better than current experimental limits for WIMPs with masses $m_\chi \gtrsim 100\,$GeV. 

Significant improvements with respect to these projected sensitivities could be made by better control of the systematic error of the neutron-induced background, or by using materials with lower concentration of radioactive materials such as $^{238}$U. For uranium concentrations $C^{238} \lesssim 1\,$ppt (parts per trillion) in weight, paleo detectors could probe WIMP-nucleon cross sections within a factor 10 of the Xe neutrino floor indicated in Fig.~\ref{fig:SI_projection}.

%*************************
\subsection{Spin-Dependent Interactions}

\begin{figure*}
 \includegraphics[width=0.49\linewidth]{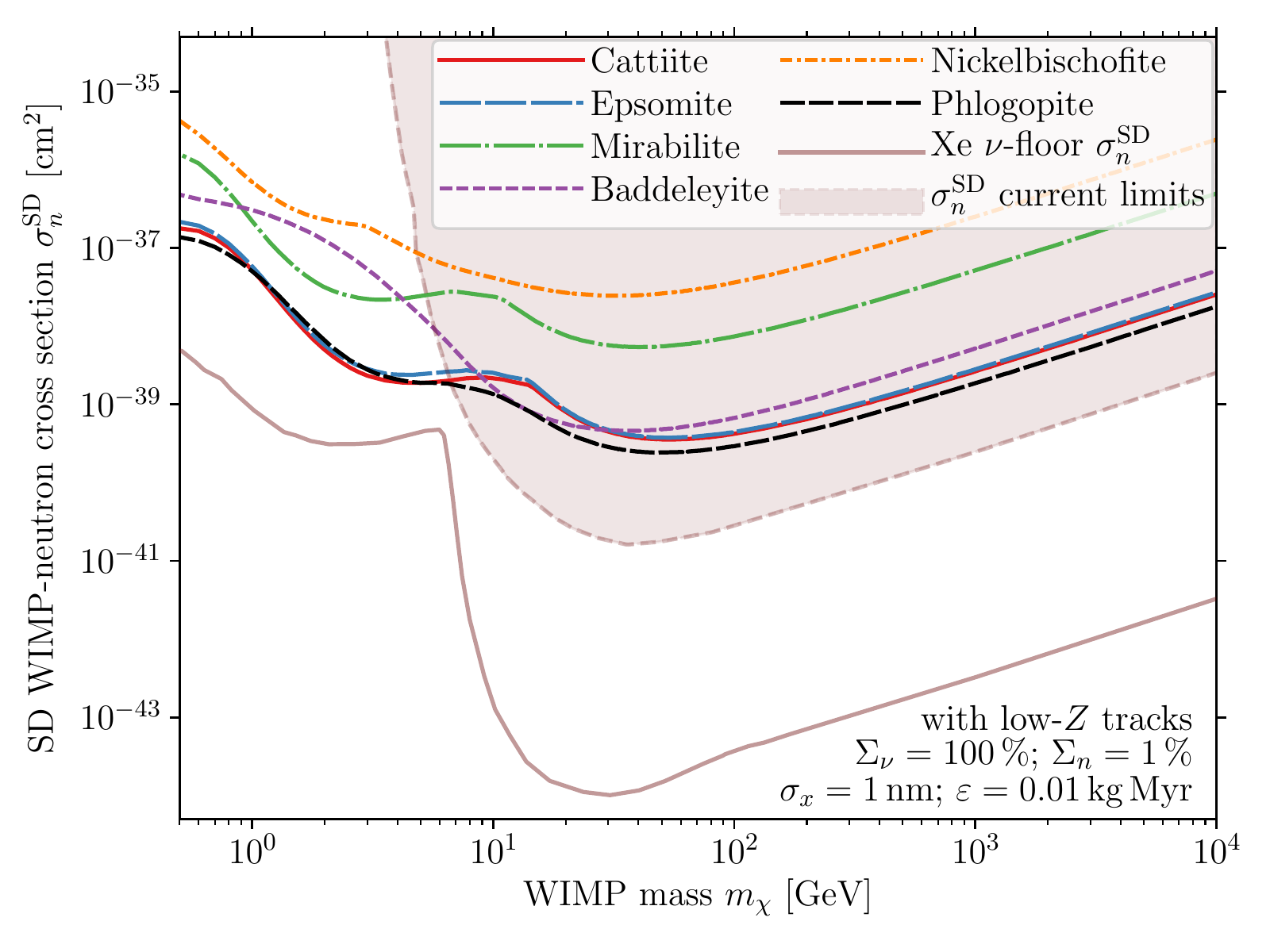}
 \includegraphics[width=0.49\linewidth]{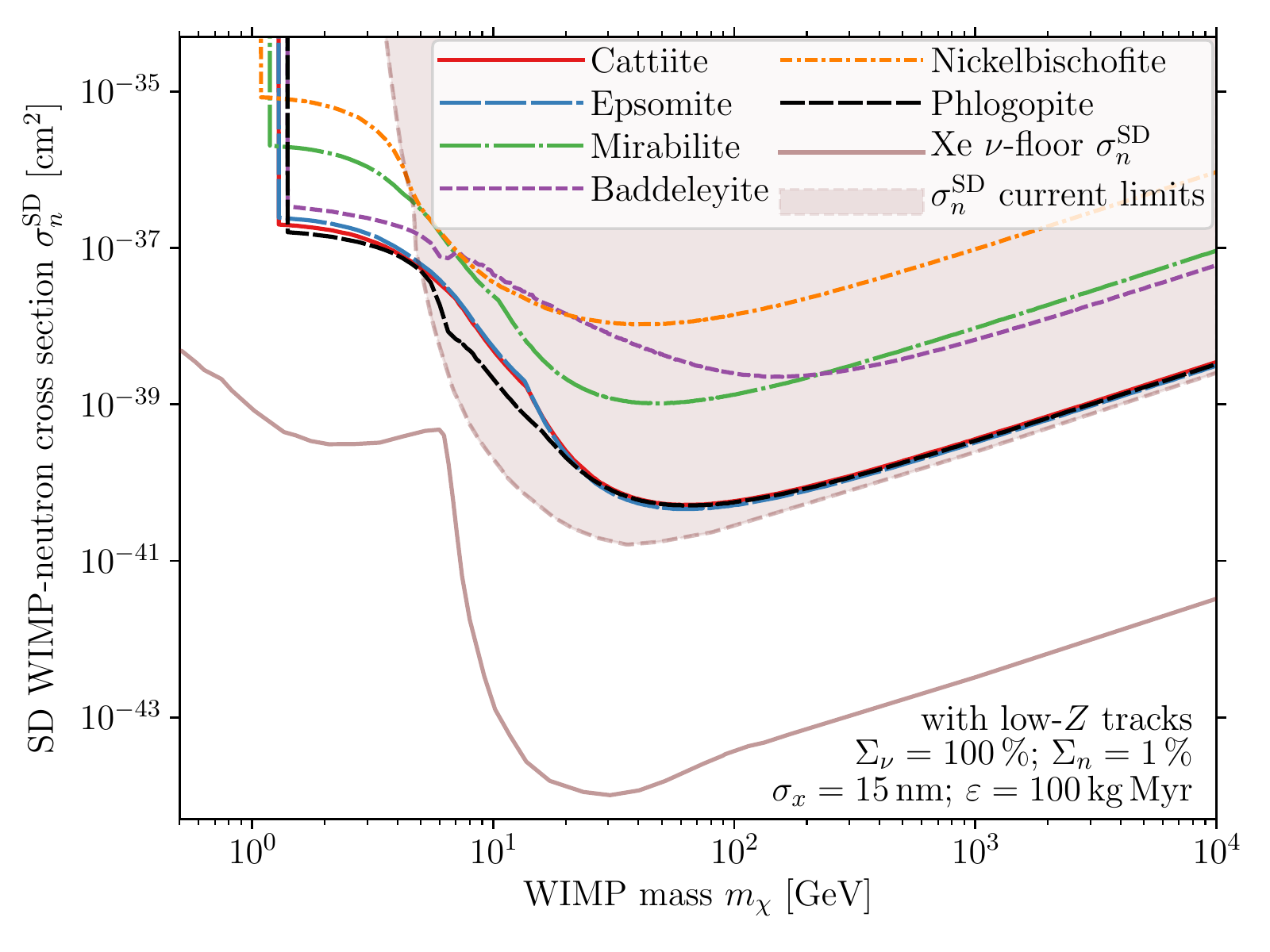}

 \includegraphics[width=0.49\linewidth]{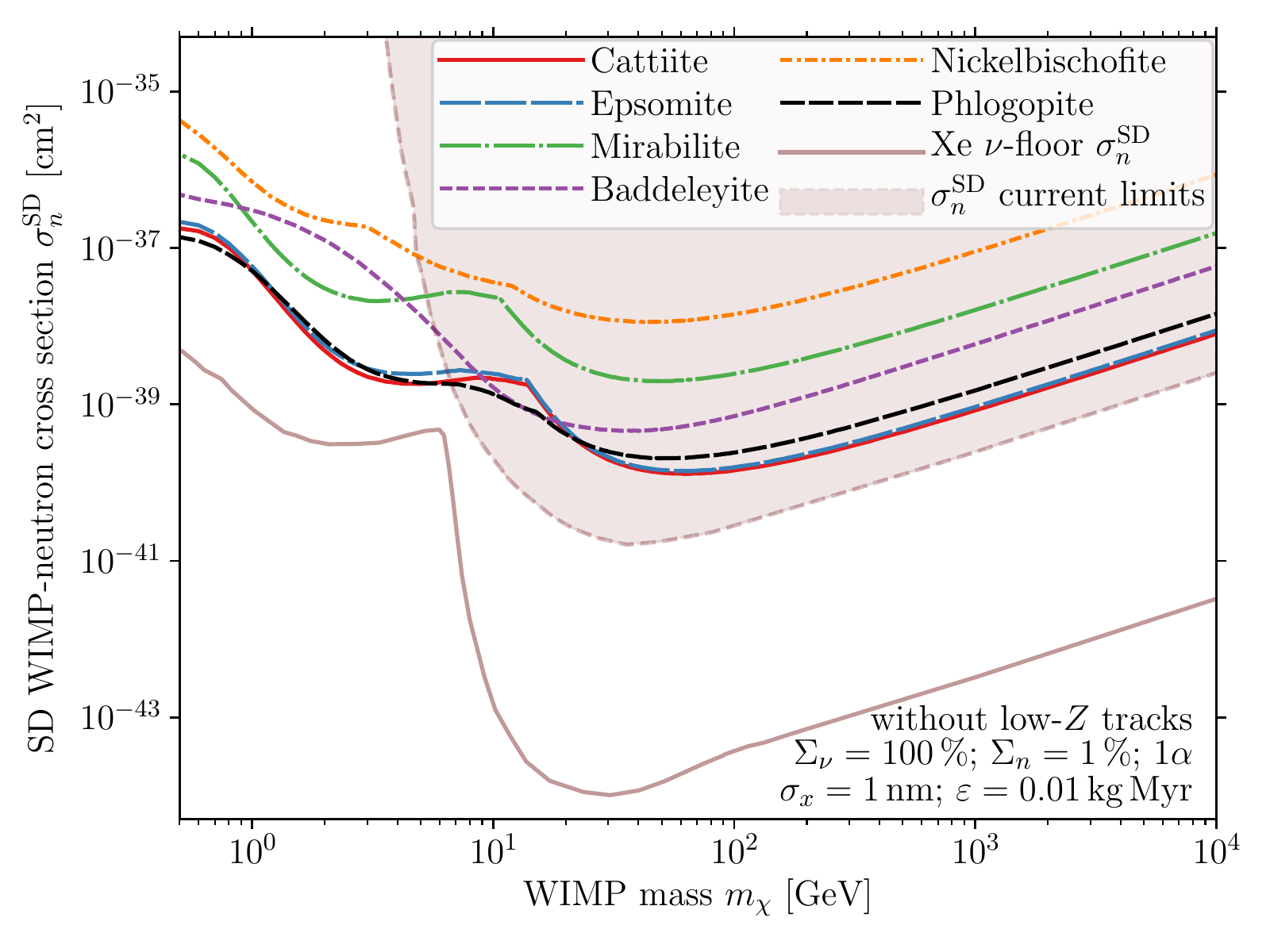}
 \includegraphics[width=0.49\linewidth]{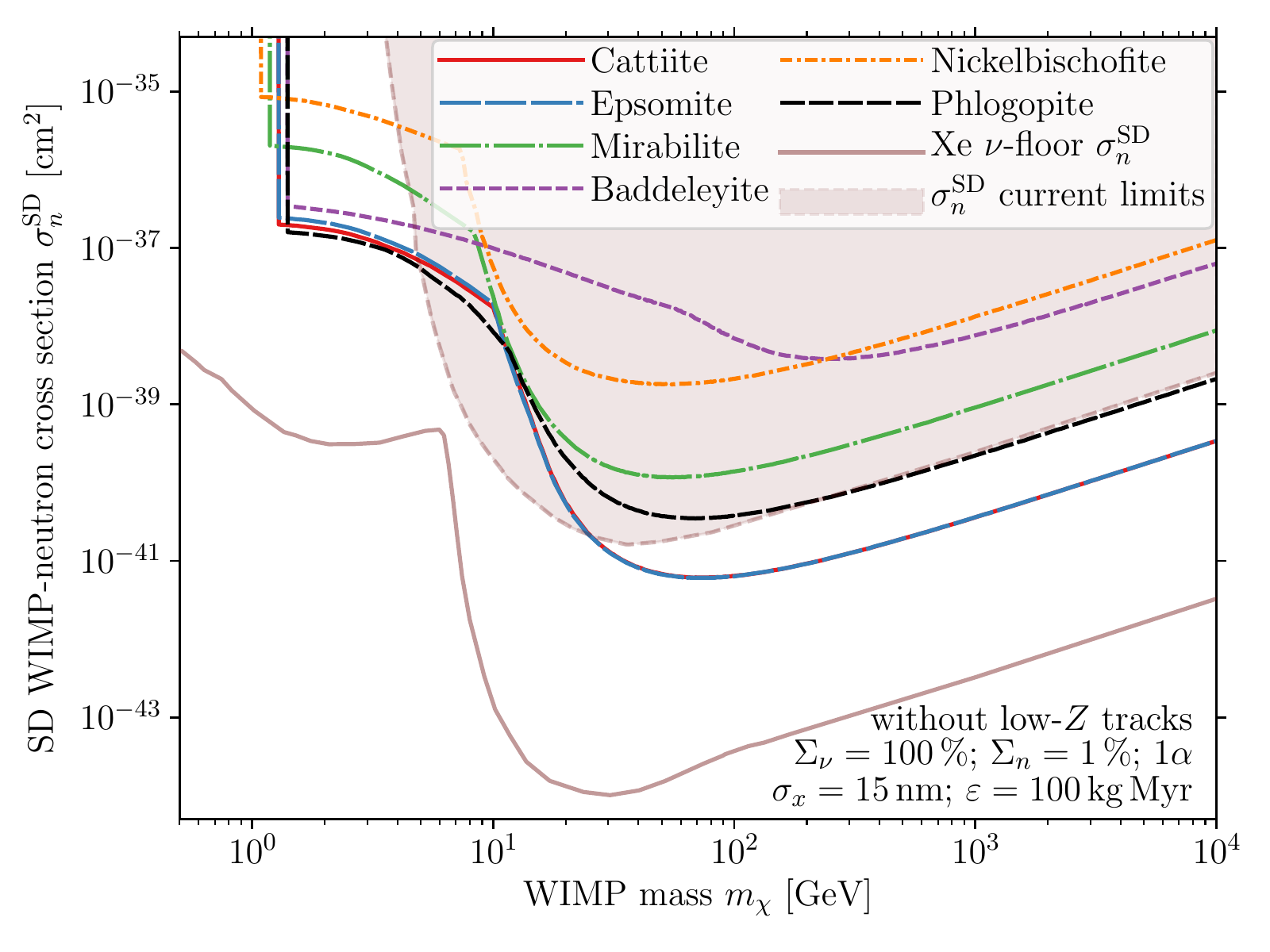}
 \caption{Same as Fig.~\ref{fig:SI_projection} but for SD scattering assuming neutron-only couplings. We show three example of MEs: Cattiite [\ccCat], Epsomite [\ccEps], and Mirabilite [\ccMir] and three examples of UBRs: Baddeleyite (\ccBad), Nickelbischofite [\ccNic], and Phlogopite [\ccPhl]. As before, we assume $^{238}$U concentrations of $C^{238} = 0.01\,$ppb in weight for MEs and $C^{238} = 0.1\,$ppb for the UBRs Baddeleyite and Nickelbischofite. For Phlogopite we assume $C^{238} = 0.01\,$ppb, since large chunks of Phlogopite with linear dimension of order $1\,$m and $C^{238} \lesssim 0.01\,$ppb have been found in natural deposits~\cite{Price:1963}. The shaded area shows current upper limits from direct detection experiments~\cite{Fu:2016ega,Akerib:2017kat}, and the light gray line indicates the neutrino-floor for SD neutron-only interactions in Xe~\cite{Ruppin:2014bra}. Note that the lines corresponding to Cattiite and Epsomite overlap for a wide range of WIMP masses in all panels shown. In the top-right panel the lines for Cattiite, Epsomite, and Phlogopite overlap for $m_\chi \gtrsim 30\,$GeV.}
 \label{fig:SDn_projection}
\end{figure*}

\begin{figure*}
 \includegraphics[width=0.49\linewidth]{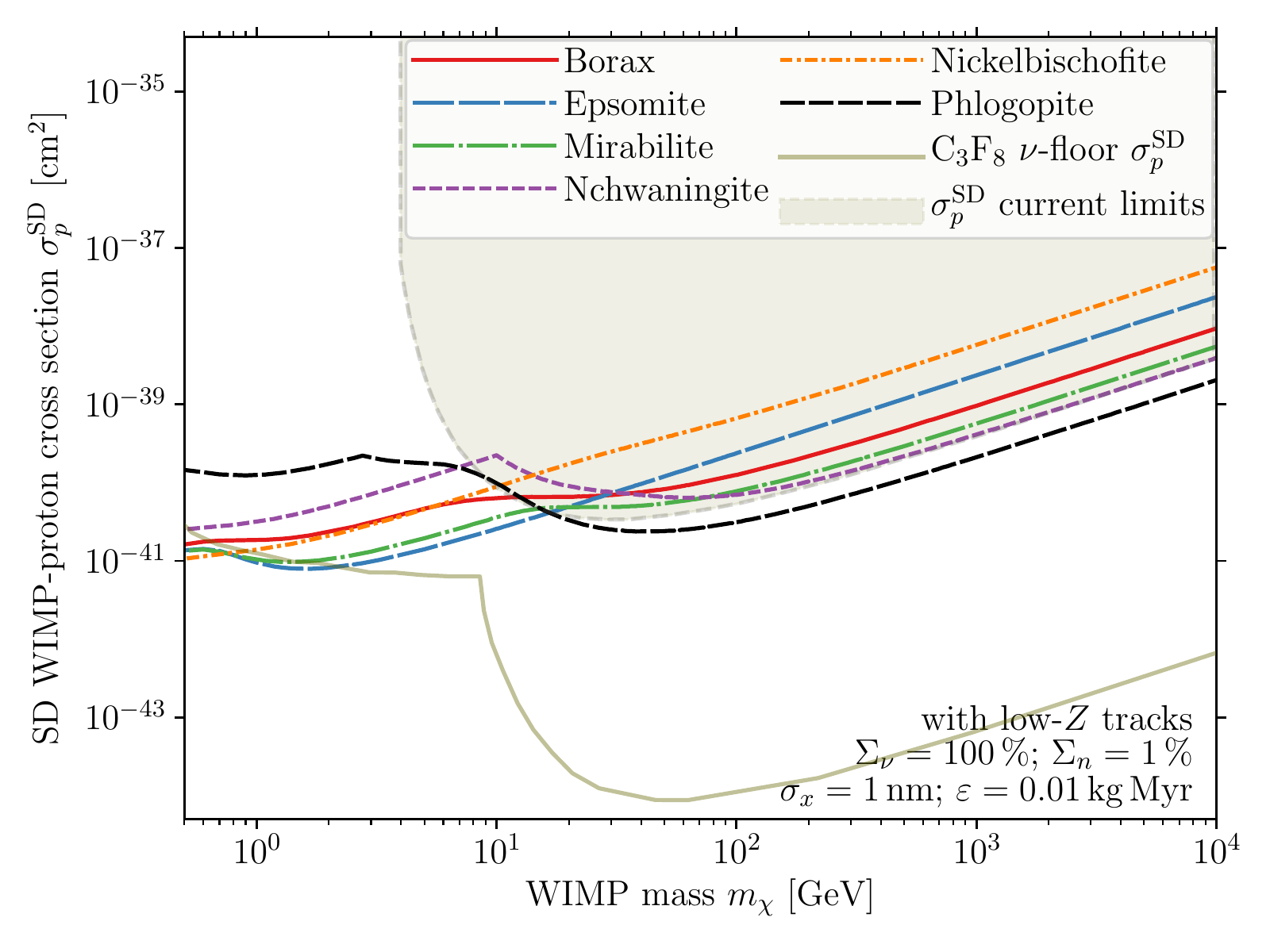}
 \includegraphics[width=0.49\linewidth]{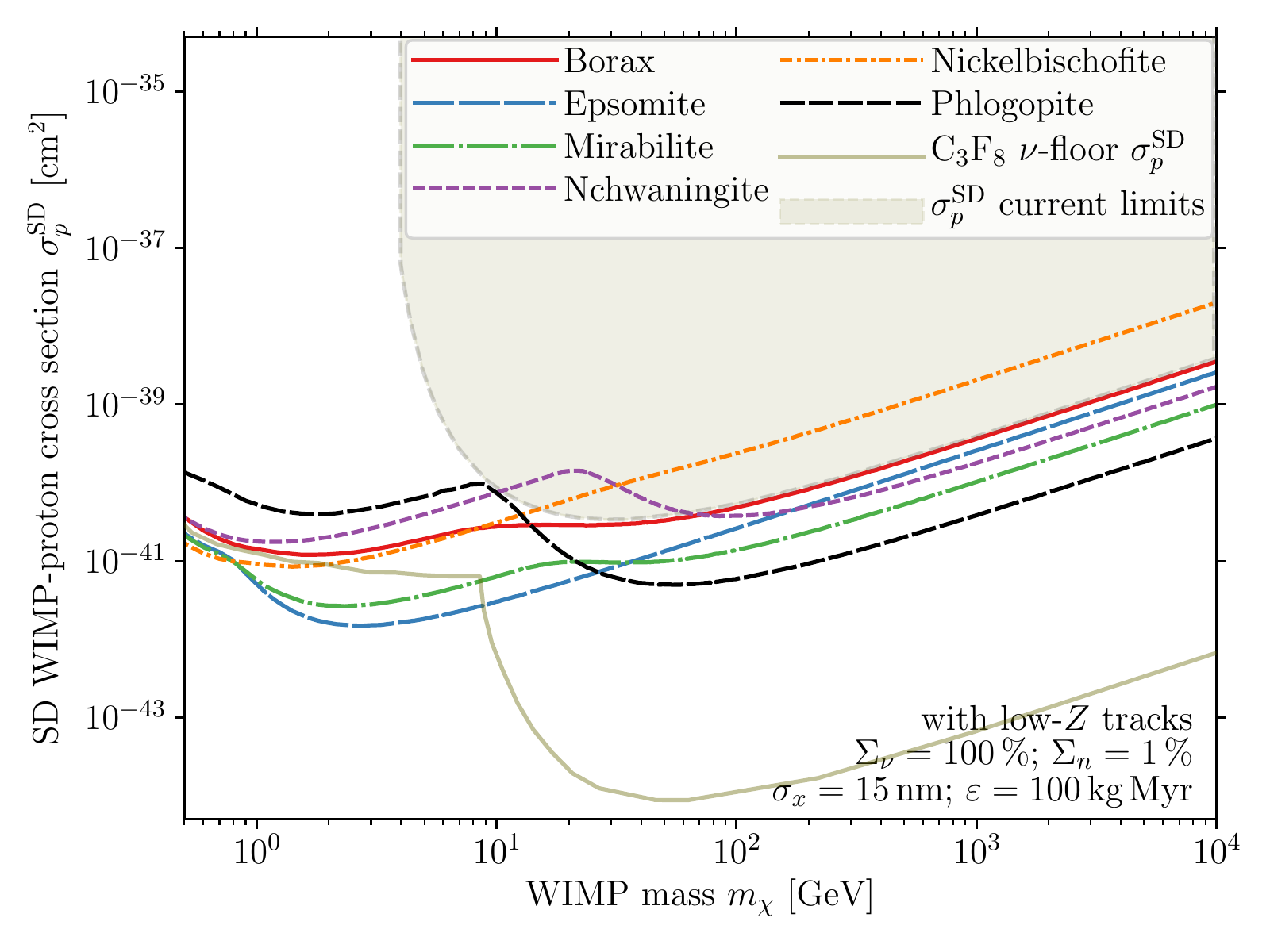}

 \includegraphics[width=0.49\linewidth]{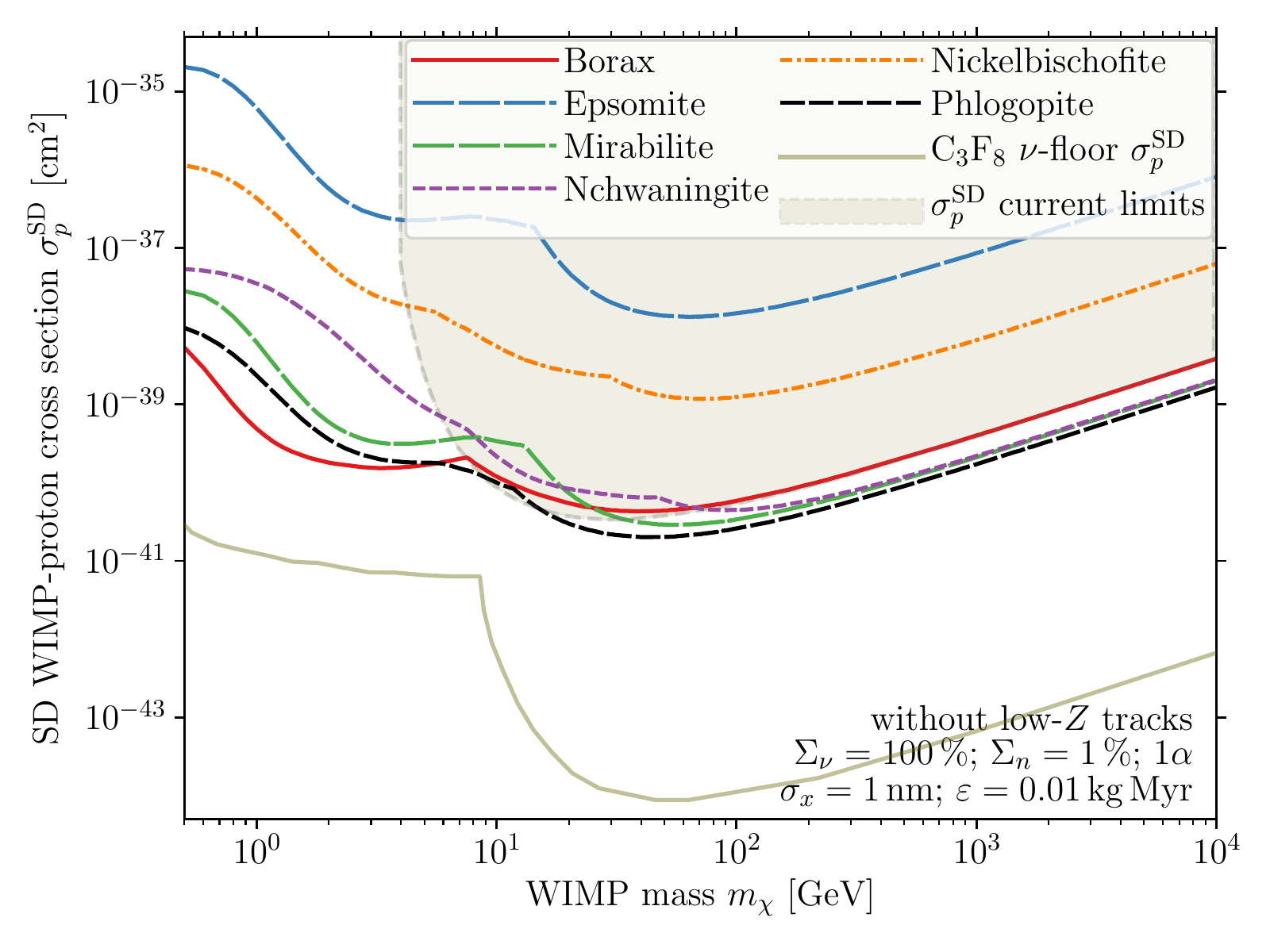}
 \includegraphics[width=0.49\linewidth]{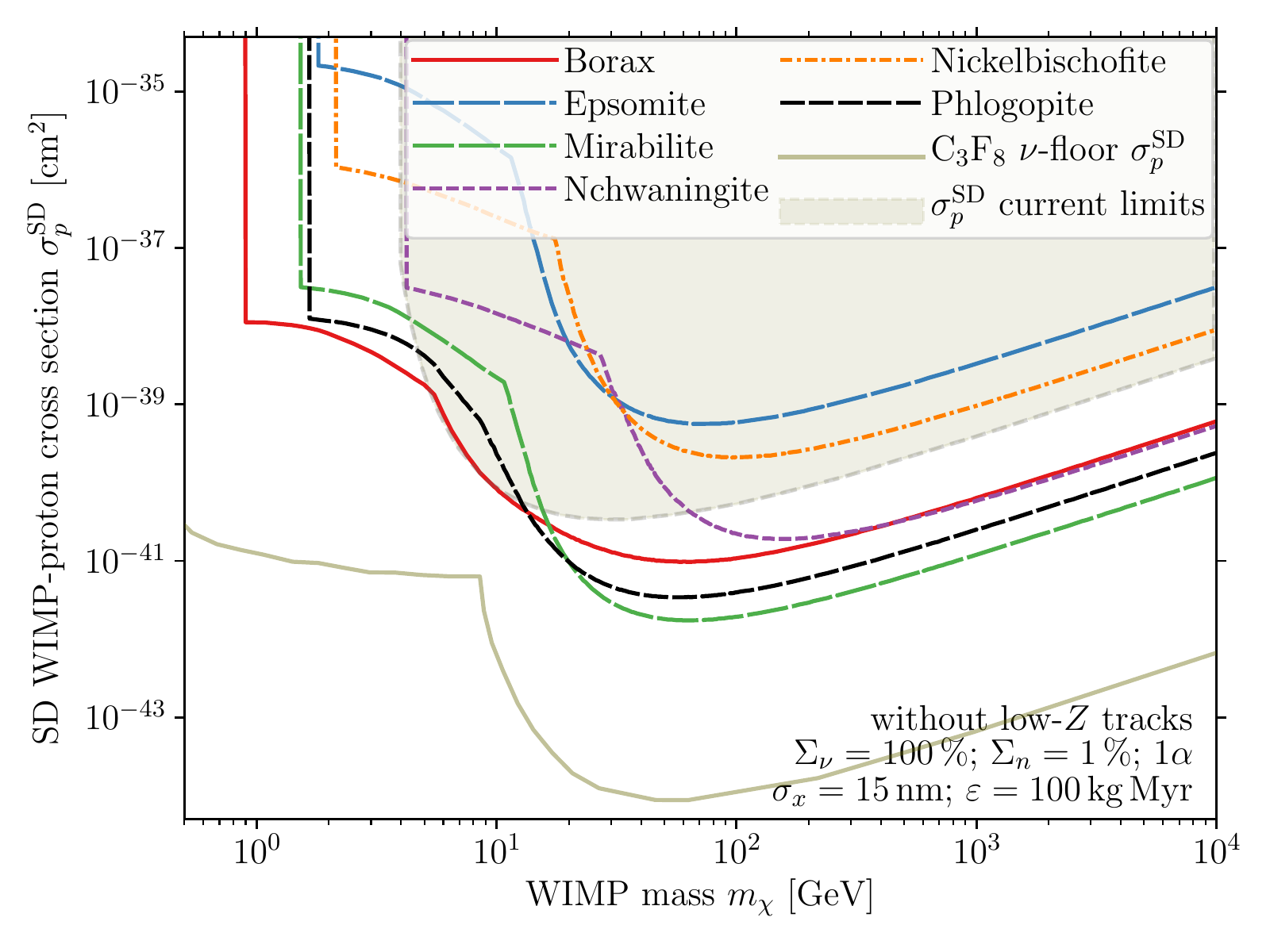}
 \caption{Same as Fig.~\ref{fig:SI_projection} but for SD scattering assuming proton-only couplings. We show three example of MEs: Borax [\ccBor], Epsomite [\ccEps], and Mirabilite [\ccMir] and three examples of UBRs: Nchwaningite [\ccNch], Nickelbischofite [\ccNic], and Phlogopite [\ccPhl]. As before, we assume $^{238}$U concentrations of $C^{238} = 0.01\,$ppb in weight for MEs and $C^{238} = 0.1\,$ppb for the UBRs Nchwaningite and Nickelbischofite. For Phlogopite we assume $C^{238} = 0.01\,$ppb, since large samples of Phlogopite with such concentrations have been found in natural deposits~\cite{Price:1963}. The shaded area shows current upper limits from direct detection experiments~\cite{Amole:2016pye,Amole:2017dex}, and the light gray line indicates the neutrino-floor for SD neutron-only interactions in C$_3$F$_8$~\cite{Ruppin:2014bra}.}
 \label{fig:SDp_projection}
\end{figure*}

We present sensitivity projections to Spin-Dependent (SD) WIMP-nucleon interactions for two different cases: neutron-only and proton-only couplings, corresponding to $a_p=0$ and $a_n=0$, respectively, in Eqs.~\eqref{eq:SSD_qd},~\eqref{eq:SSD_q0}. In principle, the sensitivity of paleo-detectors to SD interactions is quite dependent on the ratio of the effective WIMP-proton and WIMP-neutron couplings $a_p/a_n$, since typical target minerals will contain both elements sensitive to WIMP-proton and to WIMP-neutron SD interactions. We choose to present sensitivities for the neutron-only and proton-only cases here in order to allow for easy comparison to limits and projections from direct detection experiments.

\subsubsection{Neutron-Only}
In Fig.~\ref{fig:SDn_projection} we show the sensitivity to SD neutron-only couplings for the MEs Cattiite [\ccCat], Epsomite [\ccEps], and Mirabilite [\ccMir] and the UBRs Baddeleyite (\ccBad), Nickelbischofite [\ccNic], and Phlogopite [\ccPhl]. Note that as before we assume $^{238}$U concentrations of $C^{238} = 0.01\,$ppb in weight for the MEs and $C^{238} = 0.1\,$ppb for the UBRs Baddeleyite and Nickelbischofite. For Phlogopite, we assume $C^{238} = 0.01\,$ppb, since large samples of Phlogopite have been found with such $^{238}$U concentrations~\cite{Price:1963}, although typical concentrations of uranium in Phlogopite are much larger.

Comparing the different panels, we find similar behavior as for SI interactions, cf. the discussion in Sec.~\ref{sec:SIresults}. Comparing the different targets minerals we find stronger dependence on the particular chemical composition. This is because the WIMP-nucleus cross section for any particular target nucleus is no longer controlled by the number of protons and neutrons, but by the nuclear spin and how much of the spin is carried by the neutrons. Generally, it is difficult to find good target materials with much of the nuclear spin carried by the neutrons as this would require an unpaired neutron. Such isotopes are quite rare in nature, for example, $^{17}$O only makes up 0.04\,\% of natural oxygen. The most promising isotopes which are relatively common in nature appear to be $^{25}$Mg, comprising 10\,\% of natural magnesium, and $^{29}$Si, comprising 5\,\% of natural silicon. Another isotope with large nuclear spin is $^{91}$Zr, comprising 11\% of natural zirconium, however, zirconium is relatively rare in nature.

The best sensitivity out of the minerals shown in Fig.~\ref{fig:SDn_projection} is found in the MEs Cattiite and Epsomite, which both contain Mg. For lighter WIMPs, the sensitivity of Phlogopite, which contains magnesium and silicon, is comparable. For heavier WIMPs, Phlogopite is not competitive since it contains only a small fraction of hydrogen, yielding too little suppression of the neutron-induced background. For comparison, we also show the ME Mirabilite and the UBRs Baddeleyite and Nickelbischofite. Mirabilite and Nickelbischofite do not display competitive sensitivity to neutron-only interactions since they contain too few target elements with nuclear spin carried by neutrons. Baddeleyite does contain zirconium, however, it does not contain hydrogen, rendering neutron induced backgrounds too large. 

Despite the difficulties in finding good target minerals for neutron-only interactions, we can note that the sensitivity of paleo-detectors to low-mass WIMPs $m_\chi \lesssim 10\,$GeV extends to $\sigma_n^{\rm SD} \sim 10^{-38}\,$cm$^2$ for $m_\chi \sim 10\,$GeV and cross sections approximately one order of magnitude larger for $m_\chi \sim 1\,$GeV. For heavier WIMPs, paleo-detectors still promise sensitivity better than current upper limits from direct detection, but only in ME target minerals containing hydrogen and elements with sizeable fraction of the spin carried by neutrons, such as magnesium.

\subsubsection{Proton-Only}
Compared to the neutron-only case, it is much easier to find suitable target materials for SD WIMP-proton interactions. Elements with unpaired protons and sizeable nuclear spin are common in nature, e.g. H, B, F, Na, Al, K, or Mn. In Fig.~\ref{fig:SDp_projection}, we show the sensitivity for three examples of MEs, Borax [\ccBor], Epsomite [\ccEps], and Mirabilite [\ccMir], and three UBRs, Nchwaningite [\ccNch], Nickelbischofite [\ccNic], and Phlogopite [\ccPhl]. As before, we assume $^{238}$U concentrations of $C^{238} = 0.01\,$ppb in weight for the MEs Borax, Epsomite, and Mirabilite as well as for Phlogopite, and $C^{238} = 0.1\,$ppb for Nchwaningite and Nickelbischofite.

When comparing the different panels in Fig.~\ref{fig:SDp_projection}, the same arguments as for SI interactions in Sec.~\ref{sec:SIresults} apply. However, note that for proton-only SD interactions, the difference between the {\it with low-$Z$ tracks} and the {\it without low-$Z$ tracks} scenarios is particularly large. This is because hydrogen nuclei, i.e. protons, are excellent targets for SD WIMP-proton scattering of low-mass WIMPs and in addition, recoiling hydrogen nuclei have large ranges in typical target materials, cf. Fig.~\ref{fig:Nch_Range}. The resulting sensitivity in the {\it with low-$Z$ tracks} scenario extends to cross sections as small as $\sigma_p^{\rm SD} \sim 10^{-42}\,$cm$^2$ for WIMPs with masses of a few GeV. Due to the long range of hydrogen, the sensitivity to low-mass WIMPs is retained when using read-out methods with worse spatial resolution, e.g. SAXs; cf. the top right panel of Fig.~\ref{fig:SDp_projection}. The importance of the hydrogen induced tracks also explains why all target materials shown in Fig.~\ref{fig:SDp_projection} have similar sensitivity.

In the {\it without low-$Z$ tracks} scenario, the behavior is more similar to the SI and SD neutron-only cases. The best targets, Mirabilite and Phlogopite, are those containing a sizeable fraction of heavier target elements with large nuclear spins and unpaired protons in addition to hydrogen, e.g. F or Na, and having low concentrations of $^{238}$U. The sensitivity to low-mass WIMPs extends to $\sigma_p^{\rm SD} \sim 10^{-40}\,$cm$^2$ for $m_\chi \sim 10\,$GeV and approximately one order of magnitude smaller cross sections for $m_\chi \sim 1\,$GeV WIMPs. For heavier WIMPs with masses $m_\chi \gtrsim 30\,$GeV, the sensitivity is a factor $10-100$ better than current limits.

Note that as discussed for SI interactions at the end of Sec.~\ref{sec:SIresults}, the sensitivity of paleo-detectors to SD WIMP-nucleus interactions may be improved significantly if systematic uncertainties of the backgrounds are smaller than we have assumed here, or if target materials with lower concentrations of radioactive contaminants are available.

%*************************
\section{Discussion} \label{sec:Discussion}
%*************************
In this article, we have given a detailed discussion of our paleo-detector proposal. Paleo-detectors constitute a radically different approach to direct detection than conventional experiments: instead of instrumenting a (large) target mass in the laboratory and searching for WIMP-induced nuclear recoils in real time, we propose to examine ancient minerals for the traces of WIMP-nucleon interactions. This proposal rests upon the principle of solid state track detectors: in certain minerals, a recoiling nucleus will induce a damage track which, once created, persists over geological time-scales. Since paleo-detectors record WIMP-nucleon interactions over time scales as long as $\sim 1$\,Gyr, reconstructing tracks in relatively small target masses suffices to achieve exposures (the product of target mass and integration time) orders of magnitude larger than what seems feasible in conventional direct detection experiments. 

Here, we have identified two realistic, though ambitious read-out scenarios: Reconstructing tracks with helium-ion beam microscopy combined with ablation of read-out layers with pulsed lasers, target masses of $\mathcal{O}(10)\,$mg may be investigated with spatial resolutions of $\sim 1\,$nm. Translated to conventional direct detection quantities, such resolution would allow for nuclear recoil energy thresholds $\mathcal{O}(100)\,$eV, comparable to what is achieved in cryogenic bolometric detectors. Sacrificing some spatial resolution for larger target masses, we propose to read out $\mathcal{O}(100)\,$g of material with small-angle X-ray scattering, allowing for spatial resolutions of $~15\,$nm. Such spatial resolution would allow for nuclear recoil energy thresholds of $\mathcal{O}(1)\,$keV, similar to conventional direct detection experiments using liquid noble gas targets. However, the corresponding exposures are larger by a few orders of magnitude than what conventional direct detection techniques envisage in the next decades.

We have described the most relevant background sources in detail in Sec.~\ref{sec:Bkg}. One major advantage of paleo-detectors compared to conventional direct detection experiments is that the comparatively small target masses can be obtained from depths much greater than those of the underground laboratories in which conventional detectors must be operated. This makes cosmogenic backgrounds all but negligible. The dominant background sources in paleo-detectors will be nuclear recoils induced by neutrons from radioactive processes and by neutrinos. Broadly speaking, we identify two background regimes: For low-mass WIMPs $m_\chi \lesssim 10\,$GeV, the background budget is dominated by solar neutrinos. For heavier WIMPs, the dominant source of backgrounds are neutron induced nuclear recoils. In our background calculation we include both the neutrons induced by the spontaneous fission of heavy radioactive contaminants, in particular $^{238}$U, and the neutrons induced by $(\alpha,n)$-reactions of $\alpha$-particles from decays of heavy radioactive contaminants with the nuclei comprising the target material.

The selection of target minerals for paleo-detectors is heavily informed by our background study. Suitable minerals must record tracks, preserve the tracks over geological time-scales, and be as pure from radioactive contaminants as possible in order to reduce backgrounds induced by radioactivity to an acceptable level. We identify two classes of material as suitable for paleo-detectors: Minerals found in marine evaporite deposits and minerals found in ultra-basic rocks. Both type of minerals are much purer than typical materials found in the Earth's crust. When searching for heavier WIMPs where the background is dominated by neutron-induced recoils, it is advantageous to use minerals which in addition contain hydrogen. This is because hydrogen is an effective moderator of fast neutrons, thus, the presence of hydrogen significantly lowers the background induced by the neutrons from radioactivity. 

In Sec.~\ref{sec:Reach} we present the projected sensitivity of paleo-detectors for a range of target materials and for the two different read-out scenarios discussed above. While we have already presented prospects for probing canonical spin-independent WIMP-nucleus interactions in Ref.~\cite{Baum:2018tfw}, we present results here for a larger selection of minerals. In addition, we present projected sensitivities for spin-dependent WIMP-nucleus interactions in the usual proton-only and neutron-only interaction benchmark scenarios used by conventional direct detection experiments. In all cases, WIMP-nucleon cross section many orders of magnitude smaller than current experimental upper bounds can be probed for light WIMPs with masses $m_\chi \lesssim 10\,$GeV. For heavier WIMPs where the dominant background source is neutron induced recoils, the projected sensitivity strongly depends on the presence of hydrogen. For the uranium concentrations assumed in this work, WIMP-nucleon cross sections a factor of a few to $\sim 100$ smaller than current experimental upper bounds can be probed with paleo-detectors using target minerals which contain hydrogen. Note that significant improvements with respect to these projections are possible; for example, if target materials with uranium concentrations of less than approximately one part per trillion in weight are available, the background for heavier WIMPs would be dominated by nuclear recoils induced by atmospheric and supernova neutrinos. In such a situation, the sensitivity of paleo-detectors would extend to cross sections more than one order of magnitude smaller than what is shown here. The sensitivity may also be improved by using more sophisticated analysis techniques than the simple cut-and-count approach employed here, for example a spectral analysis~\cite{PAPER3}.

We would like to comment on some interesting possibilities arising from the paleo-detectors approach, which we leave for future work. For example, using a series of target materials of different ages it would be possible to obtain information of the time-variability of nuclear recoil events over scales as long as $1\,$Gyr. In the case of WIMP DM, such an approach would allow for studies of the sub-structure of the DM halo: The age of the oldest available minerals is larger than the period of rotation of the Sun around the galactic center. While sub-structure such as ultra-compact mini-halos~\cite{Berezinsky:2007qu,Ricotti:2009bs} or tidal streams~\cite{Stiff:2001dq,Freese:2003tt,Zemp:2008gw}, typically renders conventional direct detection experiments less sensitive due to a decrease in the local DM density, in paleo-detectors, the signal may be enhanced from the Earth passing through overdense DM regions. Using target materials which recorded WIMP signals over different times, it may be possible to obtain information about such sub-structure of the DM halo. 

The sensitivity and exposure time also makes paleo-detector interesting for a host of applications beyond WIMP DM searches. Examples include studying the time-variability of the fluxes of cosmic rays, or of neutrinos from the Sun or supernovae. Another example would be the study of proton decay facilitated by the large exposure.

In order to pave the way towards paleo-detectors, we plan to carry out a number of feasibility studies in the near future. We intend to use natural minerals obtained from close to the surface to demonstrate the reconstruction of fossil tracks with the read-out methods described in this work. In order to demonstrate the feasibility of paleo-detectors for DM searches, we will create signals similar to those which WIMP DM may induce by irradiating target samples with neutrons. 

Conventional direct detection experiments have been carried out for approximately three decades. Despite detectors becoming ever larger and more sophisticated, they have not delivered (conclusive) evidence of WIMP-nucleon interactions as yet. Conventional detectors will become increasingly expensive and challenging to operate. Thus, paleo-detectors are a timely proposal for an alternative strategy to extend the sensitivity of direct detection experiments to much of the remaining WIMP parameter space.

%*************************
\begin{acknowledgments}
The authors would like to thank F.~Avignone, J.~Blum, J.~Collar, J.~Conrad, D.~Eichler, R.~Ewing, T.~Edwards, A.~Ferella, A.~Goobar, B.~Kavanagh, C.~Kelso, D.~Snowden-Ifft, L.~Stodolsky, H.~Sun, K.~Sun, and C.~Weniger for useful discussions.
SB, AKD, KF, and PS acknowledge support by the Vetenskapsr\r{a}det (Swedish Research Council) through contract No. 638-2013-8993 and the Oskar Klein Centre for Cosmoparticle Physics. 
SB, KF, and PS acknowledge support from DoE grant DE-SC007859 and the LCTP at the University of Michigan. 
\end{acknowledgments}
%*************************

\bibliography{DMbib}

\end{document}